\documentclass[12pt]{article}

\usepackage{latexsym}

\font\tenmsbm=msbm10 scaled 1200
\font\sevenmsbm=msbm9
\newfam\msbmfam
\textfont\msbmfam=\tenmsbm
\scriptfont\msbmfam=\sevenmsbm

\makeatletter
\@addtoreset{equation}{section}
\makeatother

\newcommand{\eref}[1]{(\ref{#1})}

\def\beq{\begin{equation}}
\def\eeq{\end{equation}}
\def\bea{\begin{eqnarray}}
\def\eea{\end{eqnarray}}
\def\bet{\begin{tabular}}
\def\eet{\end{tabular}}
\def\pa{\partial}

\def\ve{\varepsilon}
\def\qua{\quadratello}

\catcode`@=11
\def\lsim{\mathchoice
  {\mathrel{\lower.8ex\hbox{$\displaystyle\buildrel<\over\sim$}}}
  {\mathrel{\lower.8ex\hbox{$\textstyle\buildrel<\over\sim$}}}
  {\mathrel{\lower.8ex\hbox{$\scriptstyle\buildrel<\over\sim$}}}
  {\mathrel{\lower.8ex\hbox{$\scriptscriptstyle\buildrel<\over\sim$}}} }
\def\gsim{\mathchoice
  {\mathrel{\lower.8ex\hbox{$\displaystyle\buildrel>\over\sim$}}}
  {\mathrel{\lower.8ex\hbox{$\textstyle\buildrel>\over\sim$}}}
  {\mathrel{\lower.8ex\hbox{$\scriptstyle\buildrel>\over\sim$}}}
  {\mathrel{\lower.8ex\hbox{$\scriptscriptstyle\buildrel>\over\sim$}}} }
\def\croce{\displaystyle / \kern-0.2truecm\hbox{$\backslash$}}
\def\lqua{\lower4pt\hbox{\kern5pt\hbox{$\sim$}}\raise1pt
\hbox{\kern-8pt\hbox{$<$}}~}
\def\gqua{\lower4pt\hbox{\kern5pt\hbox{$\sim$}}\raise1pt
\hbox{\kern-8pt\hbox{$>$}}~}
\def\mma{\lower1pt\hbox{\kern5pt\hbox{$\scriptstyle <$}}\raise2pt
\hbox{\kern-7pt\hbox{$\scriptstyle >$}}~}
\def\mmb{\lower1pt\hbox{\kern5pt\hbox{$\scriptstyle >$}}\raise2pt
\hbox{\kern-7pt\hbox{$\scriptstyle <$}}~}
\def\mmc{\lower4pt\hbox{\kern5pt\hbox{$<$}}\raise1pt
\hbox{\kern-8pt\hbox{$>$}}~}
\def\mmd{\lower4pt\hbox{\kern5pt\hbox{$>$}}\raise1pt
\hbox{\kern-8pt\hbox{$<$}}~}
\def\lsu{\raise4pt\hbox{\kern5pt\hbox{$\sim$}}\lower1pt
\hbox{\kern-8pt\hbox{$<$}}~}
\def\gsu{\raise4pt\hbox{\kern5pt\hbox{$\sim$}}\lower1pt
\hbox{\kern-8pt\hbox{$>$}}~}
\def\croce{\displaystyle / \kern-0.2truecm\hbox{$\backslash$}}
\def\ali{\hbox{A \kern-.9em\raise1.7ex\hbox{$\scriptstyle \circ$}}}
\def\2frecce{\hbox{\lower 0.3ex\hbox{$\leftarrow$} 
\hbox{\kern-1.3em\raise 0.3ex\hbox{$\rightarrow$}}}}
\def\quad@rato#1#2{{\vcenter{\vbox{
        \hrule height#2pt
        \hbox{\vrule width#2pt height#1pt \kern#1pt \vrule width#2pt}
        \hrule height#2pt} }}}
\def\quadratello{\mathchoice
\quad@rato5{.5}\quad@rato5{.5}\quad@rato{3.5}{.35}\quad@rato{2.5}{.25} }

\begin{document}

\begin{titlepage}

\begin{flushright}
Preprint DFPD 99/TH/22\\
June 1999\\
\end{flushright}

\vspace{1truecm}

\begin{center}

{\Large \bf Duality--invariant Quantum Field Theories}
\par\vspace{0.4cm}
{\Large \bf of Charges and Monopoles}

\vspace{2cm}

{K. Lechner\footnote{kurt.lechner@pd.infn.it}  and
{P.A. Marchetti}\footnote{pieralberto.marchetti@pd.infn.it}} 
\vspace{2cm}

{ \it Dipartimento di Fisica, Universit\`a degli Studi di 
Padova,

\smallskip

and

\smallskip

Istituto Nazionale di Fisica Nucleare, Sezione di Padova, 

Via F. Marzolo, 8, 35131 Padova, Italia
}

\vspace{1cm}

\begin{abstract}
\vspace{0.5cm}

We present a manifestly  Lorentz-- and $SO(2)$--Duality--invariant local 
Quantum Field 
Theory of electric charges, Dirac magnetic monopoles and dyons. 
The manifest invariances 
are achieved by means of the PST--mechanism. The dynamics for classical point 
particles is described by an action functional living on a circle, if the 
Dirac--Schwinger quantization condition for electric and magnetic charges 
holds. 
The inconsistent {\it classical} Field Theory depends on an arbitrary, but 
fixed, external vector field, a generalization of 
the Dirac--string. Nevertheless, the Quantum Field Theory, obtained from 
this classical 
action via a functional integral approach, turns out to be independent of     
the particular vector field chosen, and thus consistent, if the 
Dirac--Schwinger quantization condition holds.  We provide explicit 
expressions for the generating functionals of observables, 
proving that they are Dirac--string independent. 
Since Lorentz--invariance is manifest at each step, the quantum theory 
admits also a manifestly  diffeomorphism invariant coupling to external
gravity. Relations  with 
previous formulations, and with $SO(2)$--non invariant theories are clarified.

\end{abstract}

\end{center}
\vskip 0.5truecm 
\noindent PACS: 11.15.-q, 14.80.Hv, 11.30.Cp; Keywords: quantum field theory, 
monopoles, duality.
\end{titlepage}

\newpage

\baselineskip 6 mm


\section{Introduction}

The construction of relativistic quantum field theories for electric 
charges and Dirac magnetic monopoles encountered in the past two main 
difficulties: 
the implementation of Lorentz--invariance and the realization of the
global $SO(2)$--duality group as a manifest symmetry of the theory. While
the first symmetry is a consistency requirement for the theory itself, the 
second appears as a supplementary property which is inherent to the 
classical generalized Maxwell equations, including magnetic charges.
The problem of Lorentz--invariance is usually entangled with the dependence
of the quantum theory on the unphysical Dirac--string. More precisely, in 
this paper we call ``Dirac--string" of a particle  
the two--dimensional 
hypersurface swept out by the string during its time evolution; its boundary 
is the space--time particle trajectory.
  
The fundamental starting point for any theory of charges and monopoles are
the classical generalized Maxwell equations:
\bea
\label{1a}
\pa^\mu F_{\mu\nu}&=&j_\nu^e\\
\label{1b}
\pa^\mu *F_{\mu\nu}&=&j_\nu^g,
\eea 
where $j^e$ and $j^g$ are the electric and magnetic currents, and 
$*F_{\mu\nu}={1\over 2}\varepsilon_{\mu\nu\rho\sigma}F^{\rho\sigma}$ is the
dual field strength. These equations imply current conservation but need
to be accompanied by the equations of motion for the charged matter.
For $N$ classical point--particles, dyons, with electric and magnetic charges
$(e_r,g_r)$ and masses $m_r$, $r=1,\cdots,N$, they involve the generalized 
Lorentz--force and read
\beq
\label{1c}
m_r{du_r^\mu\over d\tau_r}=\left(e_rF^{\mu\nu}(y_r)+g_r*F^{\mu\nu}(y_r)\right)
u_{r\nu},
\eeq
where $y^\mu_r$ are the particles trajectories, 
$u^\mu_r={dy^\mu_r\over d\tau_r}$, and  $\tau_r$ is the proper time.

For a classical field theory \eref{1c} has to be replaced by the Dirac
or Klein--Gordon equations, and this requires also the introduction of
vector potentials.

In spite of the manifest Lorentz--invariance of \eref{1a}--\eref{1c}, the 
construction of a classical invariant action, from which they can be 
deduced, 
involves already some problem: one has to renounce either to locality 
\cite{SCHW1,SCHW2} or to manifest Lorentz--invariance \cite{ZW1}. 
These problematic aspects  which one encounters at the very beginning
 -- specially the missing manifest Lorentz--invariance --
prevented for long time also the construction of a consistent quantum theory.
That for the 
classical point--particle theory these features can eventually be saved 
\cite{SCHW2}, relies heavily on the 
point--particle nature itself of the charged matter.

At the quantum level the implementation of Lorentz--invariance
has been achieved first by Brandt, Neri and Zwanziger
\cite{ZW2}, using a 
functional integral approach based on a classical local and manifestly  
$SO(2)$--invariant action which breaks, however, Lorentz--invariance, 
depending on a fixed constant four--vector $n^\mu$. Despite of the explicit
breaking of the Lorentz group in this approach, the classical action for 
point particles
turns out to be Lorentz--invariant if it is defined on a circle, and if the 
Dirac--Schwinger quantization condition holds. On the other hand, the
classical {\it field} theory is always inconsistent in that it breaks the 
Lorentz group, while, strikingly enough, the quantum field theory based
on this inconsistent classical field theory action turns again out to be 
Lorentz--invariant, if the Dirac--Schwinger quantization condition holds.
The reason for this is essentially that the quantum
theory can be traced back, through a multi--path Feynman expansion
of the functional integral, to classical point--like trajectories (a
realization of wave/particle quantum duality), and that
the classical point particle theory is consistent and admits an invariant
exponentiated action. 

In short, there exists a consistent quantum 
field theory of charges and monopoles, if the Dirac--Schwinger quantization 
condition holds. There exist, actually, two inequivalent classes 
of consistent theories:
the theories of the first class are $SO(2)$--invariant, manifestly or not, 
while the theories of the second class are  
only invariant under the discrete duality group $Z_4$, 
despite of the $SO(2)$--invariance of the underlying classical equations
of motion, which are for both classes \eref{1a}--\eref{1c}. 
With $Z_4$ we 
mean here the discrete electromagnetic duality group generated by 
$(e_r,g_r)\rightarrow (g_r,-e_r)$.
The difference between the two types of theories lies in
the mixed quantum interactions between charges and monopoles.

For the  $SO(2)$-- and $Z_4$--theories we have 
respectively the two distinct Dirac--Schwinger quantization conditions 
\bea
\label{1}
{\scriptstyle{1\over 2}}\left(e_rg_s-e_sg_r\right)&=& 2\pi\, n_{rs} \\
\label{2}
e_rg_s &=& 2\pi\, n_{rs},
\eea
for each $r$ and $s$, where the $n_{rs}$ are integer.
The first is invariant under $SO(2)$ and the second only under $Z_4$.
None of the two conditions implies the other. Only if the stronger condition
\beq
\label{stronger}
{\scriptstyle{1\over 2}}\, e_rg_s = 2\pi\, n_{rs}
\eeq
is satisfied, then
the two theories coincide, as we will see in detail in the text. 

Which kind of theory describes the monopoles depends on 
the model one considers. For example, for the four--dimensional effective
heterotic string action \cite{SCHWARZ} \eref{1} is appropriate, while for 
the original Dirac monopole \cite{DIRAC} one has to use Dirac's original 
condition \eref{2}.
 
In this paper we concentrate mainly on the $SO(2)$--theory (elaborating 
only those details of the $Z_4$--theory through which it differs from
the $SO(2)$--theory), formulated according to the Pasti--Sorokin--Tonin (PST) 
method \cite{PST}. In addition to an $SO(2)$--doublet of vector potentials, 
this method requires a single non propagating scalar auxiliary field, 
in the following called $a$. 
The method  appears particularly suitable
for a formulation of a quantum field theory of dyons,  in that it allows 
a clear distinction between the problems of 
Lorentz--invariance and of Dirac--string dependence. The principal 
advantage of the PST--action is, in fact,  its manifest $SO(2)$-- and 
Lorentz--invariance in the case of free Maxwell equations, i.e. \eref{1a}
and \eref{1b}
with $j^e=0=j^g$. The free PST--action admits also a natural coupling to 
non vanishing currents, if they are kinematically conserved \cite{BERMED}, 
as in the case of classical point particles. In this situation the action
$S$ maintains its manifest $SO(2)$-- and Lorentz--invariance, for 
arbitrary values of the charges, but depends on generalized Dirac--strings. 
Nevertheless, if \eref{1} holds the functional $exp(iS)$ is 
string--independent and gives rise to the equations of motion 
\eref{1a}--\eref{1c}.
 
On the other hand, for the construction of a quantum field theory through a 
functional integral approach, one needs as starting point a classical 
{\it field} theory action which includes also the coupling to matter 
fields, charged spinors or scalars. Such an action, however, is always 
inconsistent. On a more fundamental level the reason for this is  
that there is no consistent set of {\it equations of motion} which describe
the coupling of gauge fields to dyonic matter fields, irrespective of the 
existence of an action which gives rise to them. The 
corresponding classical currents are not point--like, but 
spread out continuously and the local electric 
and magnetic fluxes are non--integer. As a consequence no  Dirac--Schwinger 
condition can save the theory.

The necessarily resulting inconsistency of the field theory action can be 
traded in 
several ways, e.g. Zwanziger \cite{ZW1} and Brandt et. al \cite{ZW2} 
sacrificed Lorentz--invariance keeping locality, while the authors of 
\cite{IENGO1} keep Lorentz--invariance but renounce to the description of 
the matter field dynamics through a local action.

In this paper, relying on to the PST--mechanism, we present a classical
field theory action which  
keeps formally Lorentz--invariance  and locality, but depends on an 
arbitrary, nowhere light--like, external fixed vector field $U^\mu(x)$ (see 
\eref{50}). 
The interpretation of this vector field is rather simple: its unique integral 
curve through a given space--time point determines the Dirac--string 
attached to that point. This will allow us to define Dirac--strings not
only for point--like particles but also for continuous current 
distributions. All correlators of observables in the corresponding quantum
field theory will be shown to be independent of the choice of the
vector field $U$, if \eref{1} holds, 
guaranteeing in this way the consistency of 
the quantum theory. (A completely analogous statement can be made also 
for the $Z_4$--theory.) This can be regarded as the main result of the 
paper. 

At the quantum level the particular choice 
$U^\mu(x)=n^\mu$ relates our formulation to the one in 
\cite{ZW1,ZW2}. 

The PST--method appears rather powerful in combining manifest duality
and Lorentz invariance at the classical level, and a variant of it
allows also to write manifestly Lorentz--invariant classical actions for 
self--dual  tensor fields (chiral bosons) in $4k+2$ dimensions \cite{PSTCHIR}. 
It turns also out to be compatible with all relevant symmetries, in 
particular with diffeomorphisms as a direct consequence of its manifest 
Lorentz--invariance. The reliability of the method at the quantum level 
has been examplified in \cite{KL}, where the gravitational anomalies for 
chiral bosons have been derived performing a perturbative 
one--loop analysis of the PST--action. The results of the present paper 
underline, once more, that the method is perfectly well suited also at the
quantum level, and that it allows a direct control of 
Lorentz--invariance in quantized theories. 

For the reasons just explained, in the PST--approach the coupling of the 
classical field theory of dyons to external gravity, represented by a 
classical metric $g_{\mu\nu}(x)$, can be achieved simply through the minimal 
prescription, at least in a topologically trivial manifold. 
In the presence of gravity the classical action depends again on a given 
vector field $U^\mu(x)$.
In the corresponding quantum theory, where $g_{\mu\nu}$ and $U^\mu$ remain
classical fields, under a change of $U\rightarrow U'$ the correlators
of the observables change, as in the flat case, by a phase factor, the 
``Dirac--anomaly"; but, 
since this phase factor is a topological term it is metric independent,
and since for the flat theory it equals unity under \eref{1}, it equals 
unity also for the theory in a curved background. 

The analysis of the quantum field theory in the present paper is made 
at a non--perturbative (although formal) level, using a functional integral 
approach, but
ignores ultraviolet divergences; some regulator is always 
implicitly assumed. For a discussion of the renormalizability of the 
quantum theory of dyons, in different approaches, in particular of the 
stability under renormalization of \eref{1} and \eref{2}, we refer the 
reader to \cite{IENGO2,BN}.

In the next section we give a self--contained account of the classical 
PST--approach for the $SO(2)$--theory, using the concise 
formalism of differential forms, which we will keep 
throughout this paper. Section three is devoted to the classical point
particle theory and establishes the Dirac--Schwinger condition as a 
necessary and sufficient condition for a consistent classical action 
principle. In section four, which is a kind of intermezzo, 
we give a short account of the $Z_4$--theory;
the formulation of this theory does not necessitate the PST--techniques and
requires only the introduction of a single vector potential.
In section five we return to the $SO(2)$--theory and present its
unconsistent classical field theory version, while in section six we
prove the consistency of its quantum 
field theory version, paying particular attention to the correlation 
functions of observables.  Section seven is devoted to a discussion of 
duality symmetries, $\vartheta$--angles, 
and the relation between the PST--approach and previous approaches to 
the theory of interacting dyons. Section eight contains some concluding 
remarks. Some basic identities for differential forms and a technical proof 
are relegated to the appendix.

\section{The PST--approach}

In this section we give a self--contained account of the PST--method,
applied to Maxwell's equations. First we introduce our basic notations
concerning differential forms and present an equivalent form for \eref{1a}
and \eref{1b}, suitable for the PST approach.

The components of a $p$--form $\Phi_p$ are defined by
$$
\Phi_p={1\over p!}\,dx^{\mu_1}\cdots dx^{\mu_p}\,\Phi_{\mu_p\cdots \mu_1}.
$$
The Hodge dual of $\Phi_p$, a $(4-p)$--form, has components
$$
(*\Phi_p)_{\mu_1\cdots\mu_{p-4}}={1\over p!}\,
\varepsilon_{\mu_1\cdots\mu_{p-4}\nu_1\cdots \nu_p}\,
\Phi^{\nu_1\cdots \nu_p},
$$
and the differential $d$ acts from the right, $d^2=0$. On $p$--forms the Hodge
duality operator squares to $*^2=(-1)^{p+1}$.
Forms in the image of $d$ are called exact and forms in the kernel of $d$ are 
called closed. Since we work in $R^4$ all closed forms are also exact.
The product between forms will
always be an exterior (wedge) product and the wedge symbol $\wedge$ will be 
omitted. 

In the language of forms  Maxwell's equations are
written in terms of the field strength two--form $F$ and of the 
three--forms $(J^e,J^g)$, which are the Hodge duals of the currents 
$(j^e,j^g)$, as 
\bea
\nonumber
d*F&=&J^e\equiv J^1 \\
dF&=&- J^g \equiv J^2.
\label{3}
\eea
These equations imply current conservation, which reads now
$$
dJ^1=0=dJ^2,
$$
and these, in turn,  allow the introduction of a doublet of two--forms 
$C^I$, the ``strings", satisfying
\beq
\label{4}
J^I=dC^I, \ \ I=1,2.
\eeq
In the sequel with the capital indices $I,J,K,\ldots$ we will indicate 
$SO(2)$--doublets. The two--forms $C^I$ are defined only modulo
exact 2--forms, i.e. we have invariance under
\beq
\label{4a}
C^I\rightarrow C^I + dH^I.
\eeq
For the time being we make an arbitrary but fixed choice of $C^I$ and we 
anticipate only that this ambiguity will correspond precisely to a Dirac--string
change, as it will become clear in the next section.

Introducing a doublet of one--form gauge potentials,
$A^I=dx^\mu A^I_\mu$, Maxwell's equations for conserved external currents
can be put in the equivalent manifestly $SO(2)$-- and $SO(1,3)$--invariant
form
\bea
\label{5}
F^I&=&dA^I+C^I \\
\label{6}
F^I&=&*\varepsilon^{IJ}F^J.
\eea
$\ve^{IJ}$ is the $SO(2)$--invariant antisymmetric tensor,
with $\ve^{12}=+1$, and in the following we will suppress the 
$SO(2)$--indices, e.g. \eref{6} will be written simply as $F=*\ve\,F$. In 
particular, in bilinears a contraction of those indices is always 
understood. 
The equivalence of \eref{5}, \eref{6} with \eref{3} is established 
via \eref{4} and through the 
identification $F=F^2$. The eq. \eref{5} has to be viewed as a definition
of $F^I$ while the pseudo self--duality relation \eref{6} produces really the 
dynamics.

The invariance of the curvature doublet $F$ under ``string
changes", requires that \eref{4a} has to be accompanied by
\beq
\label{4b}
A^I\rightarrow A^I - H^I. 
\eeq 

The PST--approach allows now to write a manifestly invariant action
\cite{PST,BERMED} for the pseudo self--duality equation of motion \eref{6}. One 
introduces a single  auxiliary
scalar field $a(x)$ and the corresponding one--form
$$
v={da\over \sqrt{-\pa_\rho a\, \pa ^\rho a}}\equiv dx^\mu v_\mu,
$$
whose components satisfy $v^2=v^\mu v_\mu=-1$. It is also convenient
to introduce the associated vector--field, which we indicate with the same
symbol, $v=v^\mu\pa_\mu$, and to indicate with  $i_v$ its  interior product
with a $p$--form.

The PST--action can then be written as the integral of a four--form,
\beq
\label{pst}
S_0[A,C,a]={1\over 2}\int F\,{\cal P}(v)\,F + dA \,\ve\,C.
\eeq
${\cal P}(v)$ is a symmetric operator which acts in the space of 
two--forms and on the $SO(2)$--indices as
$$
{\cal P}^{IJ}(v)=vi_v*\delta^{IJ} +\left(vi_v-{1\over 2}\right)\ve^{IJ}.
$$
The operators $vi_v$ and $*vi_v*$ map $p$--forms in $p$--forms and
project a form respectively on its components along $v$ and orthogonal
to $v$; their basic properties are given in the appendix.
 
In the compact notation used in \eref{pst} the usual Maxwell action 
with a single 
$F_{\mu\nu}$, $-{1\over 4}\int d^4x F^{\mu\nu}F_{\mu\nu}$, 
reads ${1\over 2}\int 
F*F$. 

The second term in \eref{pst} can also be written as
\beq
\label{WZ}
{1\over 2}\int dA \,\ve\,C=-{1\over 2}\int A \,\ve\,J=
{1\over 2}\int d^4x\, A_\mu \,\ve\,J^\mu. 
\eeq
It is a kind of Wess--Zumino term
and represents the standard ``electrical" couplings of both currents. 
Notice, however, that the couplings carry a factor of one--half w.r.t. ordinary
Q.E.D. This discrepancy is balanced by the fact that both currents carry
also a ``magnetic" interaction, appearing  in the first term in the form
$dA+C$. In this $SO(2)$--symmetric formulation there is, actually, no way to 
distinguish between magnetic and electric {\it currents}, both are
electrically {\it and} magnetically coupled.

What determines, ultimately, the form of the PST--action are its 
particular local symmetries. Under generic variations of the fields $a$ and
$A$ one has, in fact
\beq 
\label{7}
\delta S_0=\int 
{1\over 2}\,d\left({1\over \sqrt{-(\pa a)^2}}\,v\,f 
\,\ve\,f\right)\delta a
- d\left(vf\right)\,\ve\,\delta A,
\eeq
where we defined the doublet of one--forms
$$
f=i_v(F-*\ve F).
$$
From \eref{7} one sees that the action is invariant under the following
transformations:
\bea
\label{8}\delta A&=& d\Lambda \\
\label{10}\delta A&=& \Phi\, da\\
\label{9}\delta A&=& - {\varphi\over \sqrt{-(\pa a)^2}}\,f, \quad 
            \delta a= \varphi.
\eea
The transformation parameters are the doublets of scalars $\Lambda$ and 
$\Phi$, and the single scalar $\varphi$. The transformations
in \eref{8} are just ordinary $U(1)$ gauge transformations for the vector
potentials $A$, and \eref{9} states that the field $a$ is a non propagating
auxiliary field which can be shifted to any value.   
The transformations \eref{10} allow to reduce the second order equation
of motion for the gauge fields to the first order pseudo self--duality 
relation 
\eref{6}. To see this we write the equations of motion for $A$ and $a$,
which can be read directly from \eref{7},
\bea
\label{11}
d\,(vf)&=&0\\
\label{12}
d\left({v\over \sqrt{-(\pa a)^2}}\,f 
\,\ve\,f\right)&=&0.
\eea
The equation of motion for $a$ \eref{12} is a consequence of the 
$A$--equation \eref{11} and
the general solution of the latter is $vf=da\,d{\widetilde{\Phi}}$, for some 
scalar doublet ${\widetilde{\Phi}}$; through a transformation \eref{10}, 
with $\Phi={\widetilde{\Phi}}$, one can set this doublet to zero and obtains
$f=0 \leftrightarrow F=*\ve\,F$, which is the desired result.

In the sequel an important role will be plaid by the ``Dirac--anomaly"
of the PST--action. It is defined as the (finite) variation of the action
under string changes \eref{4a},\eref{4b}. The first term in \eref{pst}
is invariant by construction while the Wess--Zumino term contributes
with
\beq
\label{DA}
A_D[H]\equiv \delta S_0= {1\over 2}\int C \,\ve\,dH= 
                            {1\over 2}\int J \,\ve\,H. 
\eeq
Despite of this anomaly the equations of motion for the
gauge--fields are invariant under string changes, as we saw, but this 
anomaly will be no longer harmless when we couple the PST--action to 
{\it dynamical} matter, as we will see in the next section.

\subsection{The electromagnetic field strength}
  
An issue left open by the PST--approach is the correct 
identification of the electromagnetic field--strength tensor. If the
equations of motion \eref{11} are satisfied {\it and} the symmetries in 
\eref{10} have been fixed as above, then the field strength is clearly say
$F=F^2$, and  $F^1=*F$. But these tensors can not be identified with the 
electromagnetic field strength off--shell, since they are
not invariant under \eref{10} and \eref{9}, even if \eref{11} holds. 
The correct off--shell 
electromagnetic tensor is represented by the two--form doublet 
\beq
\label{13}
K^I\equiv F^I-vf^I,
\eeq
which is uniquely characterized by the following symmetry properties:
\par
1) $\delta K=0$ under \eref{8} and \eref{10} and under string changes.
\par
2) $K$ is invariant under \eref{9} on--shell (in fact, $\delta K =
{\varphi\over \sqrt{-(\pa a)^2}} (1 + *\ve)\, i_v\, d(v f)$). 
\par
3) $K=*\ve\,K$.
\par
4) $K$ reduces on--shell to $F$, if the symmetry \eref{10} is fixed 
          as above.
\par\noindent
In particular, the $K$'s are invariant under all symmetries on--shell 
and this is just what is required for an observable.  
For example, the quantum correlators of the field strength tensors in a 
path integral representation are obtained by insertion of the fields $K$.
The pseudo self--duality property 3) 
ensures then the ``uniqueness" of these 
correlators as they involve only a single field
strength, say $K^1$, and its dual; if one had used the $F's$ instead of the 
$K's$ 
then the correlators for $F^1$ would differ from the ones of $*F^2$. 

Notice also that the equations of motion for $A$ \eref{11} can be written 
alternatively as
$$
dK=dC.
$$

\subsection{The effective action}

We conclude this section with the derivation of the effective 
current--current action associated to the PST--action, which is obtained
upon performing the functional integration over the gauge fields $A$ and 
over the auxiliary field $a$ in \eref{pst}. These integrations require
an appropriate gauge fixing of the local symmetries \eref{8}--\eref{9}.
Since $a$ is a ``pure gauge" field, the integration over $a$ becomes trivial 
since the insertion of a $\delta$--function $\delta(a(x)-a^{(0)}(x))$ just 
works. Notice that, due to the appearance of $\pa_\mu a$ in the 
denominators of the classical action, the choice $a^{(0)}={\rm const.}$ 
is not allowed.
Minimal choices are $a^{(0)}=x^\mu n_\mu$, for some constant vector $n$,
which lead to $v_\mu^{(0)}={n_\mu\over \sqrt{-n^2}}$. We make here a generic 
choice for $a^{(0)}$ which leads to a nowhere singular $v^{(0)}$. 

To integrate over $A$ one
has to fix the symmetries \eref{8} and \eref{10}. To avoid explicit gauge
fixings, which make the derivation rather cumbersome for a generic $a^{(0)}$, 
we proceed in the following  alternative way. 
We rewrite \eref{pst} as
\beq
\label{ident}
S_0[A,C,a]=\Gamma_0[C]+{1\over 2}\int (G+D)\,\Omega(v) (G+D), 
\eeq
where $G$ and $D$ are single two--forms defined by
\bea
G&=&dA^1-*dA^2\\
D&=&C^1-*C^2.
\eea
$\Omega(v)$ is an operator, mapping two--forms in two--forms, given
by
$$
\Omega(v) = {d*d\over\quadratello}-v*v,
$$
where $\quadratello=\pa_\mu\pa^\mu$ is the D'Alambertian. 
We recall the standard
Hodge decomposition on $p$--forms:
$$
\quadratello=d\,\delta+\delta\,d,
$$
where $\delta$ is the co--differential,
$$
\delta=*d*.
$$
$\Gamma_0$ is defined by
\beq
\label{14}
\Gamma_0[C]= -{1\over 2}\int\left(dC\,{*\over\quadratello}\,dC
             -dC\,{\ve\over\quadratello}\,\delta C\right), 
\eeq
where contraction of the $SO(2)$--indices of the $C^I$ is understood.
The effective action is defined by
\bea
e^{i\Gamma[C]}&=&{\int \{ {\cal D} A\}_{gf}\,\{ {\cal D}a\}\,\delta(a-a^{(0)})
               \,e^{iS_0[A,C,a]}  
                \over
                \int \{ {\cal D} A\}_{gf}\, \{ {\cal D}a\}\,\delta(a-a^{(0)})
               \,e^{iS_0[A,0,a]} } \\
&=&e^{i\Gamma_0[C]}\,\cdot{\int \{ {\cal D} A\}_{gf}\,e^{{i\over 2}
              \int(G+D)\,\Omega(v^{(0)})*(G+D)}
              \over
               \int \{ {\cal D} A\}_{gf}\,e^{{i\over 2}
              \int G\,\Omega(v^{(0)})*G} }.
\eea
Since there exist linear gauge fixings for the symmetries
\eref{8} and \eref{10}, as shown below,
in the numerator one can now perform the
shift of integration variables
\bea 
A^1 &\rightarrow& A^1 -{\delta\over\quadratello}\,D\\
A^2 &\rightarrow& A^2 +*\,{d\over\quadratello}\,D,
\eea
proving that the numerator is $C$--independent and equals the denominator, 
so that $\Gamma[C]=\Gamma_0[C]$. 

For gauge--fixings of the kind $a^{(0)}=x^\mu n_\mu$ the functional integral 
over $A$ 
can be performed also with more conventional techniques, since in this case
the kinetic term for the gauge fields in the PST--action becomes local.
We have now $v_\mu={n_\mu\over \sqrt{-n^2}}=$const., and 
the symmetries \eref{8} and \eref{10} can be fixed imposing 
\beq
i_vA=0=d*A.
\eeq
These gauge fixings are suitable also for a generic $a^{(0)}$.
The kinetic term for the gauge fields in \eref{pst} reduces with these
gauge fixings to
\beq
{1\over 2}\int A*{\cal K}(v)\,A=-
{1\over 2}\int A*\left(\begin{array}{c}\qua+\pa_v^2 \\-T\pa_v\end{array}
                       \begin{array}{cc}T\pa_v\\ \qua+\pa_v^2
 \end{array}\right)A,
\eeq
where $T=*\,v\,d$ is an operator which sends one--forms into one--forms
and $\pa_v=v^\mu\pa_\mu$. The functional integral over $A$ in \eref{pst}
is Gaussian and one needs
only to compute the propagator--matrix for the gauge fields. Using the 
algebraic identity
$$
T^3=-(\qua+\pa_v^2)\,T,
$$
which holds for constant $v$, this matrix can be evaluated as
\beq
\label{prop}
{\cal K}^{-1}(v)=
-{1\over \qua\,(\qua+\pa_v^2)}\cdot
\left(
\begin{array}{c}\qua-{T^2\pa_v^2\over \qua+\pa_v^2}\\T\pa_v\end{array}  
\begin{array}{cc}-T\pa_v \\ \qua-{T^2\pa_v^2\over \qua+\pa_v^2}\end{array}
\right).
\eeq
The gaussian integral results then in
$$
\Gamma[C]={1\over 2}\int \left(d\,\left({\cal P}(v)+{1\over 2}\ve\right)
            C\right) {\cal K}^{-1}(v)*
  d\,\left({\cal P}(v)+{1\over 2}\ve\right)C +C\,{\cal P}(v)\,C.
$$
It is a mere exercise to show that the $v$--dependence drops out  from
this expression and that the latter coincides with \eref{14}. The 
independence of the effective action of the particular choice 
one makes for $a^{(0)}$ is clearly a consequence of the symmetry \eref{9} 
of the PST--action. 

The first term of the effective action corresponds to the usual 
diagonal current--current interaction (remember that $J=dC$), while the
second term, corresponding to a mixed electric--magnetic interaction,
involves not only $J$ but exhibits also an explicit dependence on the 
strings $C$. It is this mixed interaction which is responsible for
the Dirac--anomaly associated to $\Gamma_0[C]$. 
The effective action carries, indeed,  the same Dirac--anomaly as the 
PST--action itself: under $C\rightarrow C+dH$ one has
$$
\label{15} \Gamma_0 \rightarrow \Gamma_0+A_D[H].
$$
This simple result will play a fundamental role in establishing 
the consistency of the quantum field theory. 

\section{Point--like particles}

Classical point--like dyons are characterized by their mass $m_r$ and by
their electric and magnetic charges, $e^I_r=(e_r,-g_r)$, and sweep out 
a space--time trajectory $\gamma_r$, parametrized by arbitrary 
parameters $s_r$, $\{y^\mu_r(s_r), -\infty \leq s_r \leq \infty\}$. The 
extension of the PST--action to dynamical classical particles requires
the addition of the standard term $-\sum_{r=1}^N m_r
\int_{\gamma_r}d\tau_r$, but needs also an appropriate definition
of the two--forms $C$ in terms of the particles trajectories. For this 
purpose we make now a short digression on Poincar\`e--duality
and de Rham--currents.

\subsection{Poincar\`e--Duality and Dirac--strings}
   
Until now we did not specify the regularity properties of the differential
$p$--forms introduced, and we assumed implicitly that their
components are smooth functions. For our purposes it is, however, more
convenient to consider them as ``(de Rham) $p$--currents", i.e. linear 
functionals on the space of smooth $(4-p)$--forms 
with compact support, which are 
continuous in the sense of distributions \cite{deRham}. 
In other words, $p$--currents are
$p$--forms with distribution--valued components; we will call them 
still ``forms". 

In this space we can extend Poincar\`e duality to a map PD 
which associates to every $p$--dimensional hypersurface
$\Sigma_p$ a $(4-p)$--form 
$\Phi_{\Sigma_p}$ according to
\bea
\nonumber
{\rm PD}:\, &&\Sigma_p \rightarrow  \Phi_{\Sigma_p} \\
\int_{\Sigma_p}\Psi_p=&&\int_{R^4}\Psi_p \Phi_{\Sigma_p},
\label{16}
\eea
for any  $p$--form $\Psi_p$. This  map 
respects in particular the exact--forms $\leftrightarrow$ 
boundary--hypersurfaces correspondence, i.e.
\bea
&&{\rm PD}:\,\Sigma_p \rightarrow \Phi_{\Sigma_p} \qquad {\rm implies} \\
&&{\rm PD}:\,\pa \Sigma_p \rightarrow d\Phi_{\Sigma_p},
\eea
where $\pa$ indicates the boundary operator \footnote{For a mathematically
precise formulation, involving chains, see \cite{deRham}.}. 

A basic consequence of PD--duality is that   
the integral of $\Phi_{\Sigma_p}$ over a generic $(4-p)$--dimensional 
surface $S_{4-p}$ is an integer, counting the 
intersections with sign of $\Sigma_p$ with $S_{4-p}$. This implies, in 
particular, that the integral over $R^4$ of a product of
two such forms is also an integer,
\beq
\label{17}
\int_{R^4} \Phi_{\Sigma_p}\Phi_{\Sigma_{4-p}}= N,
\eeq
which counts the number of intersections with sign of $\Sigma_p$ 
with $\Sigma_{4-p}$. 

Linear combinations of such $p$--forms with integer coefficients are called
{\it integer} forms (they are PD--dual to integer $(4-p)$--chains 
\cite{deRham}).
The property \eref{17} holds also for generic integer forms: the integral 
of a product of two integer forms is an integer, whenever the integral
is well defined.

In this language the closed three--forms $J$ are a weighted sum of
closed integer forms, the weights being given by the charges
\beq
\label{18}
J^I=\sum_r e^I_r\, J_r,
\eeq
where the closed integer three--forms $J_r$ are the Poincar\`e duals of 
the boundaryless curves $\gamma_r$,
\beq
\label{18a}
J_r={1\over 3!}\,dx^\rho dx^\nu dx^\mu
   \ve_{\mu\nu\rho\sigma}\int_{\gamma_r}{dy_r^\sigma
\over ds_r}\,\delta^4(x-y_r)\,ds_r.
\eeq
In this case the Wess--Zumino term  of the PST--action 
can be written also in the standard form
\beq
-{1\over 2}\int A\,\ve\,J=-{1\over 2}\sum_r\,e_r\,\ve \int_{\gamma_r} A.
\eeq

Once the currents have been determined as in \eref{18}, for the strings $C$ 
subject to $dC=J$ we choose a class of solutions given by
\beq
\label{19}
C^I=\sum_r e^I_r\, C_r,
\eeq
where the $C_r$ are integer two--forms, being Poincar\`e duals 
of two--dimensional 
hypersurfaces, extending to infinity, whose boundaries are $\gamma_r$.
This implies in particular that
$$
J_r=dC_r.
$$
If the surfaces are parametrized for example 
by $Y^\mu_r(s,u)$ with
$Y^\mu_r(s,0)=y_r^\mu(s)$, then one has the explicit expressions
\beq
\label{20}
C_r ={1\over 2}dx^\nu dx^\mu \ve_{\mu\nu\rho\sigma}
 \int_0^\infty du\int_{-\infty}^{+\infty}ds\, {dY_r^\rho\over du}\,
{dY_r^\sigma\over ds}\,\delta^4(x-Y_r).
\eeq 

The way of formalizing the Dirac--string problem presented above unifies,
actually, a  variety of treatments made in the literature, which differ
through the particular classes of chosen surfaces. In \cite{IENGO1,
IENGO2}, where only closed curves are considered, the surfaces have been 
chosen as bounded ones; in \cite{ZW1,ZW2} they have been chosen to be in 
every 
point parallel to a given constant four--vector $n^\mu$: $Y^\mu(s,u)=
y^\mu(s)+n^\mu u$.
Schwinger's method \cite{SCHW1,SCHW2}
amounts to picking a single space--time curve $\xi(u)$  which extends 
from the origin to infinity, chosen once for ever, and to attaching this 
same curve to every point of the particle trajectory: $Y^\mu(s,u)=
y^\mu(s)+\xi^\mu(u)$. Schwinger considered, indeed, also double--sheeted 
but half--weighted surfaces. These are obtained from 
the single--sheeted surfaces in \eref{20} with the replacement  
$$
\int_0^{+\infty}du \rightarrow {1\over 2}\left(\int_0^{+\infty}du
-\int_{-\infty}^0 du\right),
$$
and upon defining $\xi(-u)=-\xi(u)$. In his formulation single--sheeted
surfaces gave rise to the $Z_4$--theory and double--sheeted ones to the
$SO(2)$--theory. Since our theory is manifestly $SO(2)$--invariant one 
could equally well use one or the other type of surfaces; for definiteness 
from now on we use the type \eref{20}.

Among all possible surfaces of this type there are special classes which
will become relevant at the field theory level later on. Each of these 
classes is characterized by a nowhere light--like vector field $U^\mu(x)$
$$
U^\mu(x)U_\mu(x)\neq 0.
$$ 
This condition is necessary to ensure the validity of the associated
decomposition of the identity on $p$--forms 
\beq
1={1\over U^2}\left((-)^{p+1}Ui_U+*\,Ui_U\,*\right),\quad U\equiv dx^\mu U_\mu,
\eeq
relation which allows to decompose a $p$--form uniquely in a $p$--form  
parallel to $U$ and one orthogonal to $U$. 

The strings associated to $U$ are determined requiring that they are made 
out of the integral curves of $U$ which end on the particles trajectories.
The related $C$--fields are uniquely determined through
the equations
\bea
\nonumber
dC&=&J\\
\label{21}
i_UC&=&0,
\eea
and through the boundary condition
\beq
\label{22}
C\rightarrow 0\qquad {\rm for}\qquad \, x_U\rightarrow  -\infty, 
\eeq
meaning that we require that $C(x)$ goes to zero if $x$ goes to $-\infty$
along the integral curves of the vector field $U$. 

To prove 
the uniqueness, and existence, of the solution of this system we observe that
there exists always a diffeomorphism which maps the vector field $U^\mu(x)$ 
to a constant vector $N^\mu$. Since the system \eref{21}, \eref{22}
is diffeomorphism invariant and metric independent, in the new coordinate
system it becomes
\bea
\nonumber
dC&=&J\\
\nonumber
i_NC&=&0\\
\label{23}
C&\rightarrow& 0\qquad {\rm for} \qquad x_N \rightarrow-\infty, 
\eea
where $x_N^\mu=u N^\mu $ is the coordinate along $N$ and the limit
above means $u\rightarrow -\infty$. This system has now the solution
\beq
\label{24}
C={1\over \pa_N}\,i_N\,J, 
\eeq
as can be seen observing that 
\beq
\label{25}
i_Nd+di_N=\pa_N.
\eeq
To satisfy the boundary condition in \eref{23} the inverse operator 
${1\over \pa_N}$ has to be defined by the Kernel 
\beq
\label{25a}
G(x)=\Theta(x_N)\delta^3\,(\vec x_N^\bot),\qquad \pa_N G(x)=\delta^4(x),
\eeq
where $\vec x_N^\bot$ are the three coordinates orthogonal to $x_N$ and 
$\Theta$ is the step--function. The 
uniqueness of the solution can be inferred considering the associated
homogeneous equations, which lead, thanks to \eref{25}, to $\pa_N C=0$
and this, due to the boundary condition, implies $C=0$. The solution of the
original system can be obtained from \eref{24} upon performing the inverse
diffeomorphism. 

This proof of the uniqueness of the solution for $C$ applies equally well
to integer and smooth $J$. 
  
If the currents are given as in \eref{18} and \eref{18a},
one can also exhibit an explicit expression for the unique
solution. For each curve 
$\gamma_r$ one determines the integral curves of $U$ ending on $\gamma_r$,
\bea
\nonumber
{d\over du}Y^\mu_r(s,u)&=&U^\mu(Y_r(s,u)),\quad 0\leq u\leq \infty \\
  Y^\mu_r(s,0)&=&y^\mu_r(s),
\label{25b}
\eea
and one uses the associated surfaces in the general formula \eref{20},
to  obtain 
\beq
\label{26}
C_r ={1\over 2}\,dx^\nu dx^\mu \ve_{\mu\nu\rho\sigma}\,U^\rho(x)
 \int_0^\infty du\int_{-\infty}^{+\infty}ds\, 
{dY_r^\sigma\over ds}\,\delta^4(x-Y_r).
\eeq
The total $C$'s are given by a weighted sum according to \eref{19}; one 
recognizes in particular that $i_UC=0$. It is worthwhile to notice that
the representations for $J$ and $C$ given above apply equally well when the
curves are closed; closed curves will indeed play a central role in the
quantum field theory.

Once the $C$--fields are weighted integer two--forms,
what we called previously string changes amounts now just to a change
of the surface whose boundary is a given curve: 
\beq
\label{27}
C_r\rightarrow C_r +dH_r,
\eeq
where $H_r$ is the integer one--form associated, via PD, to the 
three--dimensional hypersurface which is bounded by the old and the new 
surfaces. For the two--forms $C$ the transformations \eref{27} lead to
\bea
\label{28}
C^I&\rightarrow& C^I+dH^I,\quad {\rm with} \\
\nonumber
H^I&=&\sum_r e^I_r\,H_r.
\eea

\subsection{Action principle for classical dyons}

Now we can present an action principle for classical 
dyons. The action which describes the interaction of dyons with the 
electromagnetic field would  be given in the PST--approach by 
\beq
\label{29}
S[A,C_r,a]=S_0[A,C,a]-\sum_{r=1}^N m_r\int_{\gamma_r}d\tau_r,
\eeq
where the $C_r$--fields are defined  as above in terms of arbitrary surfaces;
thus the dynamical variables of this action are {\it formally} the strings 
$C_r$ and not the curves $\gamma_r$. But the string dependence of $S$ is 
measured by the Dirac--anomaly. Under \eref{28}, together with 
$A^I \rightarrow A^I- H^I$, one has
$$
S \rightarrow S  + A_D[H],
$$
where now
\beq
\label{30}
A_D[H]={1\over 2}\int J \,\ve\,H={1\over 2}\sum_{r,s}\,(e_r\,\ve\, e_s)
\int J_r H_s.
\eeq
The last integral is integer, because the integrand is  a product of  
integer forms, and 
${1\over 2} (e_r\,\ve\, e_s)={1\over 2}(e_r^1e_s^2-e_r^2e_2^1)$ is an integer
multiple of $2\pi$ due to the Dirac--Schwinger condition \eref{1}. 
In conclusion, for the point particle theory, the Dirac--anomaly is an 
integer multiple of $2\pi$. This suggests to consider as action functional 
the exponentiated action $e^{iS}$, which is string--independent
\footnote{The classical theory could be based also on the exponential of
a scaled action, $e^{i\lambda S}$, which would imply a quantization 
condition where ${1\over 2} (e_r\,\ve\, e_s)$ is an integer multiple of $2\pi$
modulo an overall constant; it is quantum mechanics which eventually forces
$\lambda=\hbar=1$.}. 

This would
be the right answer were it not for the presence of ``exceptional" 
configurations where a string $C_s$ intersects the boundary of another 
string $C_r$, i.e. the curve $\gamma_r$ 
\footnote{For regular curves the ``dimension"  
of the set of these intersection points is $-1$.}. 
For such configurations the 
Dirac--anomaly becomes indeed ill-defined since in the integral
$\int J_rH_s$ the intersection of $\gamma_r$ with the three--volume 
associated to $H_s$ occurs at the boundary of this three--volume. These are
precisely the configurations which were forbidden by Dirac's veto
\cite{Diracveto}, and which were counted by Schwinger \cite{SCHW2} with
weight 1/2 (instead of 0 or 1), leading him to postulate momentarily 
the stronger quantization condition ${1\over 2} (e_r\,\ve\, e_s)=4\pi n$, 
a request which he eventually abandoned. 

To cope also with these configurations
we modify the action \eref{29} as follows. For each configuration 
 $\{\gamma_r\}$, in addition to the arbitrary strings $C_r$, we introduce a set 
of ``regular" strings $\widetilde C_r$, $d\widetilde C_r=J_r$, which never 
intersect a curve 
$\gamma_s$ (this is always possible if the curves themselves do not 
intersect) and define $\widetilde C^I=\sum_r e^I_r\, \widetilde C_r$. 
Then we modify the action according to 
\bea
\widetilde S[A,C_r,\widetilde C_r,a]&=& S[A,C_r,a]+{1\over 2} \int C\,\ve\,
                 \widetilde C\\
        &=& {1\over2}\int F\,{\cal P}(v)\,F+F\,\ve\,\widetilde C
          -\sum_r m_r\int_{\gamma_r}d\tau_r,
\eea
and treat $C_r$ and $\widetilde C_r$ as dynamical variables. The added term
is ${1\over 2}\sum_{r,s}(e_r\ve e_s)\int C_r\widetilde C_s$ and, since the 
integrand is a product of integer forms, upon exponentiation it contributes
only if the $C_r$ are in exceptional configurations, i.e. if they
intersect a curve $\gamma_s$ which is, indeed, the boundary of 
$\widetilde C_s$. Notice in particular that,
since the added term is $A$--independent, $\widetilde S$ possesses still
the PST--symmetries \eref{8}--\eref{9} and leads to the same $A$--equations
of motion. 

Since $C$ appears in $\widetilde S$ only in the combination $F=dA+C$,
the modified action is now strictly invariant under a 
$C$--string change, even if \eref{1} does not hold. All the string--dependence
has now been shifted to the regular strings $\widetilde C$. Under 
$\widetilde C\rightarrow \widetilde C+d\widetilde H$ one has
$$ 
\widetilde S\rightarrow \widetilde S+ A_D[\widetilde H],
$$ 
and $A_D[\widetilde H]$ is always an integer multiple of $2\pi$. 

The action principle can be based on the functional $exp(i\widetilde S)$ 
in a 
canonical manner. One introduces a real parameter $\alpha$ and deformed
strings $C_\alpha$, $\widetilde C_\alpha$ such that $C_\alpha \rightarrow
C$, $\widetilde C_\alpha \rightarrow \widetilde C$ for $\alpha 
\rightarrow 0$, 
and one evaluates the corresponding action $\widetilde S_\alpha$. These 
deformations include in particular deformations of the curves.
The equations
of motion for the particles (Lorentz--force law) are then obtained as
\beq
\label{31}
\left({d\over d\alpha}\,e^{i\widetilde S_\alpha}\right)\Bigg|_{\alpha=0}
=0.
\eeq
As $exp(i\widetilde S)$ is invariant under string changes {\it and}
under 
the PST--symmetries so are the equations of motion; this implies
already that the Lorentz--force can depend on $C$, $A$ and $a$ only through
the tensors $K$, defined in the preceding section, and on $u^\mu_r
={dy^\mu_r\over d\tau_r}$. Apart from a
term proportional to the $A$--equations, 
\beq
\label{32}
dK^I=J^I,
\eeq
an explicit evaluation of \eref{31} leads indeed to
\beq
\label{33}
m_r{du_r^\mu\over d\tau_r}=\left(e_r\,\ve\,K^{\mu\nu}\right)(y_r)
                        \, u_{r\nu}.
\eeq
Remembering that one has also
\beq
K^I=*\ve^{IJ}\,K^J,
\label{34}
\eeq
and that $K$ is an on--shell completely invariant tensor, the system
\eref{32}--\eref{34} becomes a closed system for all observables in 
that it 
determines uniquely $K^I$ and $y_r$, once the initial conditions have
been fixed. From this point of view $K$ is truly identified with the
electromagnetic field strength and, a posteriori, there is no need to 
introduce vector potentials or strings. Alternatively, one can use  
the gauge--fixing of the PST--symmetry \eref{10} presented in the
preceding section, which leads to $K^I=F^I$ and $F^1=*F^2$, and the system
given above reduces to the system \eref{1a}--\eref{1c}, with $F=dA+C$.  

This concludes the presentation of the extension of the PST--method to a 
system of classical dynamical dyons. Summarizing we can say that the 
functional $exp(i\widetilde S)$, which is manifestly $SO(1,3)$ and
$SO(2)$--invariant, gives rise to the correct equations of motion
\eref{1a}--\eref{1c} if the Dirac--Schwinger quantization condition
\eref{1} holds. 

The next issue would be the classical field theory, 
but before addressing this point we present now the essential features 
of the $Z_4$--theory.

\section{The $Z_4$--theory}

Let us first of all stress that this theory is equivalent to the 
$SO(2)$--theory for what concerns the classical  
equations of motion, which are again \eref{1a}--\eref{1c}. It differs
from the latter only through the classical action, on which these
equations are based, and through the quantization condition, now \eref{2}, 
which ensures the consistency of the corresponding action principle. 
Another important difference lies in the fact that the classical PST--action 
is manifestly $SO(2)$--invariant, whether \eref{1} holds or not, while
for the  $Z_4$--theory $Z_4$ becomes a symmetry group of the action
only  at the quantum level, and requires \eref{2}.

\subsection{The classical and effective actions}

Since the $Z_4$ theory is not $SO(2)$-invariant we abandon in this section 
the doublet notation and introduce a single vector potential $A=A^2$
and the single curvature two--form $F=F^2=dA+C^2$. The two--forms $C^1$ and
$C^2$ maintain the meaning of the preceding section. $J^1=dC^1$ represents
now the electric current and $J^2=dC^2$ the magnetic one. The action,
analogous to $S_0$, is
\beq
\label{35}
I_0[A,J^1,C^2]=\int {1\over 2}\,F*F-dA\,C^1
              =\int {1\over 2}\,F*F+A\,J^1,
\eeq
where the electric current carries now  only an electric coupling, with
weight 1, and
the magnetic current only a magnetic one. The equation of motion for
$A$ and the Bianchi identity for $F$ reproduce the generalized Maxwell 
equations \eref{3}.

The action \eref{35} is invariant under ordinary $U(1)$ gauge transformations,
$A\rightarrow A+d \Lambda$, while under string changes 
\bea
\nonumber
C^1&\rightarrow& C^1+dH^1\\
\nonumber
C^2&\rightarrow& C^2+dH^2\\
A&\rightarrow& A-H^2,
\label{36}
\eea
it varies as
\beq
\label{37}
A_D[H^2] \equiv\delta I_0 = \int J^1 H^2.  
\eeq
Notice that under $C^1\rightarrow C^1+dH^1,\,A\rightarrow A$ the action  
is invariant. 

The quantum field theory associated to \eref{35} can be constructed
according to the recipe reported in section six. The resulting theory,
based on a functional integral, can be considered as a formal continuum
limit of the compact Q.E.D. with scalar matter fields, formulated on a 
lattice.  
In fact, due to the invariance \eref{36} we can 
view $A$ {\it mod} $H^2$ as a continuum analogue of the compact--$U(1)$ 
gauge field on the lattice. 

The current--current effective action associated to $I_0$ is obtained upon
performing the functional integration over the vector potential $A$
\beq
e^{i\Gamma^{Z_4}_0[C]}={\int \{ {\cal D}A\}_{gf}\, e^{iI_0[A,J^1,C^2]}\over
                         \int \{ {\cal D}A\}_{gf}\, e^{iI_0[A,0,0]} },
\eeq
where one needs only to fix the ordinary $U(1)$ gauge invariance. The
integral is Gaussian and gives 
$$
\Gamma^{Z_4}_0[C]=\int -{1\over 2}\left(dC^1\,{*\over \qua}\,dC^1+
dC^2\,{*\over \qua}\,dC^2\right)+dC^1\,{1\over \qua}\,\delta C^2,
$$ 
which depends on $C^1$ only through $J^1$, as does $I_0$, and carries
the same Dirac--anomaly \eref{37} as the classical action.  

The action principle for classical point--like particles
can be formulated according to the strategy adopted for the 
$SO(2)$--theory. One introduces general strings  $C_r$ and
regular strings $\widetilde C_r$ and constructs the total $C$'s and
$\widetilde C$'s as in the preceding section. The appropriate
action for dynamical dyons is then 
\beq
\widetilde I[A,\widetilde C^1,C^2]= 
\int{1\over 2}F*F-F \widetilde C^1
-\sum_r m_r\int_{\gamma_r}d\tau_r,
\eeq
and transforms under string changes, see \eref{37}, as
\beq
\delta \widetilde I= \int J^1 \widetilde H^2
= \sum_{r,s} \,e_r^1\, e_s^2 \int J_r \,\widetilde H_s. 
\eeq
The last integral is integer and the Dirac--anomaly becomes 
an integer multiple of $2\pi$, if Dirac's original quantization condition 
\eref{2} holds. The added term, $-\int C^2\widetilde C^1$, contributes
in the exponential again only for exceptional configurations, thanks to \eref{2}.
The action principle is based on $exp(i\widetilde I)$ and the 
Lorentz--force equation is again \eref{1c}. 

\subsection{Comparison with the $SO(2)$--theory}

To compare the two theories, which at the classical level are
equivalent, we compare the effective 
current--current interactions, which trigger the quantum dynamics,
and assume, for consistency, that \eref{1} {\it and} \eref{2} are satisfied.

The diagonal interactions are identical, but
the difference in the mixed interactions leads to
\beq
\label{38}
\Delta \Gamma_0\equiv\Gamma^{Z_4}_0-\Gamma_0={1\over 2}\int C^1C^2.
\eeq
For arbitrary $C$--fields $\Delta \Gamma_0$ is non vanishing. If 
the strings are weighted integer forms, as in the point  particle theory,
we have
$$
\Delta \Gamma_0 ={1\over 2}\sum_{r,s}\,e_r^1e_s^2\int C_r\,C_s
               = {1\over 2}\sum_{r,s}\,e_r^1e_s^2\, N_{rs}, 
$$
where $N_{rs}$ is a symmetric matrix of integers\footnote{For the 
self--interaction of the 
$r$--th particle a regularization is understood, which keeps $N_{rr}$ 
integer.}. What matters in the quantum field theory is eventually the 
exponential of the effective action, therefore a necessary and sufficient
condition for the identification of the two theories is 
$$
e^{i\Delta \Gamma_0}=1.
$$
This  is equivalent to 
\beq
\label{39}
{1\over 2}\left(e^1_re^2_s+e^1_se^2_r\right)=2\pi\, n_{rs} \quad 
{\rm for}\quad r\neq s, \quad {1\over 2} e^1_re^2_r=2\pi\, n_{rr}. 
\eeq 
Taking \eref{1} and \eref{2} into account, these conditions are 
satisfied only if eq. \eref{stronger}, i.e.
$$
{1\over 2}\, e_rg_s = 2\pi\, n_{rs},
$$
holds for each $r$ and $s$. 

In conclusion we can say that the consistency conditions for the 
$SO(2)$ and $Z_4$--theories are respectively \eref{1} and \eref{2}, while
\eref{stronger}
are the supplementary conditions for the identification of $Z_4$ 
as a duality subgroup of $SO(2)$
\footnote{While \eref{stronger} implies \eref{1} and 
\eref{2}, there  are solutions to this last system of equations which do 
not satisfy \eref{stronger}.}.

\subsection{$Z_4$--symmetry}

We address now the realization of the discrete duality group $Z_4$. Its
generator acts on the $C$--fields as
\bea
C^1&\rightarrow &C^2\\
C^2&\rightarrow &-C^1.
\eea
The action $I_0$ is not invariant while the transformation of
the effective action can be read from \eref{38}. Since $\Gamma_0$ is 
manifestly invariant, we deduce 
\beq
\label{40}
\Gamma^{Z_4}_0\rightarrow \Gamma^{Z_4}_0 -\int C^1C^2.
\eeq
For integer currents, due to \eref{2}, we have that $exp(i\Gamma^{Z_4}_0)$
is indeed $Z_4$--invariant. 

To display the quantum implementation of this symmetry on the action $I_0$
one has to combine the $Z_4$ transformation with an $S$--duality 
transformation of the vector potential in the functional integral (for a 
discussion of $S$--duality in the functional integral formalism, see e.g. 
\cite{P}). 
For this purpose one introduces an auxiliary two--form $H$ and a one--form 
$A^1$, the new vector potential, and considers the action
$$
I_H=\int {1\over 2}\left(H+C^2\right)*\left(H+C^2\right)
      -H\left(dA^1 +C^1\right). 
$$ 
Functional integration over $A^1$ leads to $dH=0\Rightarrow H=dA^2$ and
one recovers $I_0$. On the other hand, integrating over $H$
$$
 \int \{ {\cal D} H\}\,e^{iI_H} \equiv e^{i\widetilde I_0},
$$
one obtains 
$$
\widetilde I_0=\int  {1\over 2}\left(dA^1+C^1\right)*\left(dA^1+C^1\right)
            +dA^1C^2 +C^1C^2.
$$
Performing now the $Z_4$--transformation on the $C$'s, accompanied 
by $A^1\rightarrow A^2$, one obtains
\beq\label{41}
\widetilde I_0 \rightarrow I_0 -\int C^1C^2,
\eeq
in agreement with the transformation of the effective action, \eref{40}.

On the other hand it may be interesting to notice that the 
transformation law \eref{40} can also be deduced directly for the classical
action $I_0$, but using a non--local transformation for the vector
potential. Under
\bea
C^1&\rightarrow &C^2\\
C^2&\rightarrow &-C^1\\
A&\rightarrow &A+{*\over \qua}\,d\left((1+*)C^2-(1-*)C^1\right)
\eea
one has $I_0\rightarrow I_0-\int C^1C^2$.

At this point an interesting question is whether the action $I_0$ can be
amended for a symmetry enhancement from $Z_4$ to
$SO(2)$ to occur. This modified action would thus  represent
a non--manifestly invariant alternative for the PST--action $S_0$.  
The suggestion for such a modification comes from  \eref{38}. It suffices
indeed to put
$$
\bar I_0= I_0 -{1\over 2} C^1C^2,
$$
which has as effective action the $SO(2)$--invariant one $\Gamma_0$. But
$\bar I_0$ possesses now also a non--local $SO(2)$--invariance, the 
corresponding infinitesimal transformations, with parameter $\varphi$,
being given by
\bea
\nonumber
\delta C^1 &=& \varphi\, C^2\\
\nonumber
\delta C^2 &=& -\varphi\, C^1\\
\delta A &=& -\varphi \,{*\over \qua}\,d\left(C^2+*C^1\right).
\label{so2}
\eea
Under these transformations $\bar I_0$ is invariant irrespectively of the
validity of any quantization condition on the charges, as is $\Gamma_0$.
This action possesses the same Dirac anomaly as the
PST--action and upon adding the same regularizing term,
${1\over 2}\int C\,\ve\,\widetilde C $, produces also the correct 
Lorentz--force--law, if the $SO(2)$--invariant quantization condition 
\eref{1} holds.
Due to its manifest invariance, however, the PST--action is preferable in
many respects. 

The action $I_0$ represents a 
generalization to arbitrary strings of the action used by Schwinger
\cite{SCHW2} to deduce the quantization conditions \eref{1} and  \eref{2}
considering respectively double--sheeted and single--sheeted strings.
For single--sheeted strings, which were always implicitly assumed by us, 
$I_0$ is not $SO(2)$--invariant, as we saw above. Keeping single--sheeted
strings it becomes $SO(2)$--invariant
upon adding the term $-{1\over 2}\int C^1C^2$. 

On the other hand, if one uses the particular double--sheeted strings 
introduced by 
Schwinger \cite{SCHW2}, this added term can be seen to vanish identically. 

\subsection{The $\vartheta$--angle}

The last issue of this section concerns the introduction of 
$\vartheta$--angles in the $Z_4$--theory. In the present context this
is a question about more general consistent couplings between the currents 
and the electromagnetic field, then those contained in $I_0$. 
Taking the quantization condition \eref{2} into account,  at present the
theory depends only on a single coupling constant (apart from the 
masses). In fact, the general
solution of \eref{2} is
\bea
\nonumber
e_r&=& e_0\,n_r\\ 
g_r&=& {2\pi\over e_0}\,m_r,
\label{42}
\eea
with $n_r,m_r$ integers, where $e_0$ is the fundamental electric charge
and represents the unique coupling constant. 

For additional couplings we
have to require $U(1)$ gauge invariance and string independence. For 
dimensional reasons the unique new term which satisfies these requirements
is $\int FF$.  
Therefore, a $\vartheta$--term can be introduced consistently in the theory 
according to 
\beq
\label{43}
I_\vartheta=I_0+{e_0^2\vartheta\over8\pi^2}\int FF.
\eeq
Notice that, contrary to what happens for $\vartheta$--terms in  
Yang--Mills theories, in the present case it is not a topological term
and hence not invariant under $\vartheta\rightarrow\vartheta +2\pi$;
we will come back to this feature in sections seven and eight.

To analyze the effect of this additional coupling we compute the new 
Maxwell equations
\bea
d*F&=&J^1-{e_0^2\,\vartheta\over4\pi^2}\,J^2\\
dF&=&J^2,
\eea
which show that the electric current has been shifted by an amount 
proportional to the magnetic current, which is indeed the expected effect of 
a $\vartheta$--term, while the magnetic current has not been touched. 
The new Lorentz--force law is again of the form \eref{1c}
but the individual charges  are now given
by
\bea
\nonumber
E_r&=&e_r+{e_0^2\,\vartheta\over4\pi^2}\,g_r\\ 
\label{44}
G_r&=& g_r, 
\eea
reflecting the shift of the total currents in the Maxwell equations.

The Dirac--anomaly, instead, does not change since the 
$\vartheta$--term is string--independent. The effective action associated
to $I_\vartheta$ reflects just the shifts noted above and is given by
\beq
\label{45}
\Gamma^{Z_4}_\vartheta=
\int -{1\over 2}\left(
\left(dC^1-{e_0^2\vartheta\over4\pi^2}dC^2\right)
\,{*\over \qua}\,
\left(dC^1-{e_0^2\vartheta\over4\pi^2}dC^2\right)
+dC^2\,{*\over \qua}\,dC^2\right)+dC^1\,{1\over \qua}\,\delta C^2.
\eeq
Notice, however, that the mixed term is unaffected by the 
$\vartheta$--angle as is the Dirac anomaly. Consequently the consistency
condition remains \eref{2}, with solution \eref{42}, and the total 
individual charges become
\bea
\nonumber
E_r&=&e_0\left(n_r+{\vartheta\over2\pi}\,m_r\right)\\
\label{46}
G_r&=&{2\pi\over e_0}\,m_r.
\eea
These charges satisfy, in turn, the relation $E_rG_s -E_sG_r =2\pi\,n_{rs}$, 
but one should
keep in mind that the fundamental equations  are \eref{2} and \eref{44}.
There are, in fact, solutions of the former relation, depending on one more
parameter, which are not taken into account by \eref{46}.

One might wonder whether $\vartheta$--terms can be introduced also in the
$SO(2)$--theory. The answer, which is encoded in the different
quantization condition \eref{1}, is that these couplings are already
present because that quantization condition allows for solutions with 
non vanishing $\vartheta$--angles (see section seven).

\section{Classical $SO(2)$--field theory}

The starting point for a quantum field theory in a functional integral
approach is a classical field theory action. Since for a theory of charges
and Dirac--monopoles there is no natural string--independent
choice for such an action, we begin by 
searching for a convenient set of equations of motion, over which we 
have a more direct control. Our guiding principles will be $U(1)$ gauge
invariance, current conservation, and manifest $SO(2)$--invariance.
For simplicity we consider here dyonic matter fields which are only minimally 
coupled to gauge fields, without additional mutual interactions.

\subsection{A set of equations of motion}

The field content is a doublet of vector potentials
$A^I$ and a finite number of bosonic or fermionic complex matter 
fields $\Phi_r$, which are  
characterized by their electric and magnetic charges $e^I_r$ and by 
their masses $m_r$. 
The $U(1)$--covariant  derivative on the $r$--th matter field is given 
by\footnote{Actually, the gauge group is $U(1)\times U(1)$.}
\beq
\label{47}
D_\mu(A) \Phi_r=\left(\pa_\mu+ie^I_r\,\ve^{IJ}A^J_\mu\right)\Phi_r.
\eeq
The equations of motion for the matter fields should already imply
current conservation by themselves; this constrains them to be the standard
ones, 
\bea
\left(D^\mu D_\mu +m_r^2\right)\varphi_r&=&0\\
\left(i\gamma^\mu D_\mu-m_r\right)\psi_r&=&0,
\eea
for bosons and fermions respectively. The individual number--currents are 
defined as usual by
$$
j^\mu_r(\Phi)=\cases 
{
i\bar\varphi_rD^\mu\varphi_r +c.c. & for bosons \cr 
&\cr
\bar \psi_r\gamma^\mu\psi_r & for fermions,\cr
}
$$
and the total electric and magnetic currents are represented by the
doublet of one--forms
$$
j^I(\Phi)=\sum_r\,e^I_r\, dx^\mu j_{\mu r}(\Phi).
$$
With $J^I(\Phi)$ we indicate the smooth three--forms which are Hodge dual to
$j^I(\Phi)$; in the bosonic case they depend, actually, also on $A$. 
Current conservation is again expressed by
$$
dJ(\Phi)=0,
$$
and we impose further the generalized Maxwell equations
$$
dF=J(\Phi),\quad  F=*\,\ve F.
$$

What is missing at
this point is the link between the fields $F$ and $A$, which can not be 
independent
variables. To construct this link we follow the strategy adopted for 
the point particle theory. Current conservation allows to introduce 
a doublet of $C$--forms satisfying $J(\Phi)=dC$ and Maxwell's equations
lead then to $F=dA+C$. But the problem arising now is that the $C$--fields 
are not 
uniquely determined, they are determined only modulo exact forms, 
corresponding to ``string changes". Moreover,
under $C\rightarrow C+dH$, $A\rightarrow A-H$ the matter equations are
not invariant meaning that, contrary to what happens in the point 
particle  
theory, in the classical field theory the equations of motion are necessarily
``string--dependent". This difference between the two theories is due to the
fact that in the point particle theory the introduction of vector potentials is
optional while in the field theory it is unavoidable.

To make a specific string choice  we have to impose one 
more condition on $C$. For this purpose we introduce a fixed external vector
field $U^\mu(x)$, with properties specified in section three, and impose
$i_UC=0$.  The set of equations of motion which  govern
the classical field theory (taking e.g. the matter
fields to be bosons) are given by
\bea
\nonumber
\left(D^\mu D_\mu +m_r^2\right)\varphi_r&=&0\\
\nonumber
F&=&dA+C\\
\nonumber
F&=&*\,\ve F\\
\nonumber
dC&=&  J(\varphi)\\
\label{48}i_UC&=&0,
\eea
with the boundary condition
\beq
\label{bound}
C\rightarrow 0\qquad {\rm for}\qquad \, x_U\rightarrow  -\infty.
\eeq
As we showed in section three the last two
equations, together with the boundary condition, determine $C$ uniquely as 
a function of  $J$, and the remaining equations become then a closed
system for the vector potentials and the matter fields.

This set of equations is $SO(2)$--invariant and formally Lorentz--invariant, 
but depends on an external vector field. In this sense it is inconsistent.
We observed already the classical field theory is necessarily inconsistent,
but the way the inconsistency shows up depends
clearly on the formulation of the theory. In the set \eref{48}, once the
auxiliary fields $C$ have been eliminated, what remains is a system of non
local equations for the remaining physical fields, due to the presence 
of $U$. On the other hand, in \cite{ZW1} Zwanziger used a local action,
amended by suitable boundary conditions, where only the
physical fields show up; but, due to the presence of a constant vector
$N^\mu$, Lorentz--invariance is explicitly broken. If the boundary 
conditions are imposed, Zwanziger's equations of motion 
become non--local and coincide with the system \eref{48}, 
for $U^\mu(x)=N^\mu$. In the next subsection we will write an action 
which gives rise to \eref{48}; but even for $U^\mu(x)=N^\mu$ 
this action will not coincide with the one given in \cite{ZW1}. 
The relation between Zwanziger's framework and ours will be explained 
in section seven.

One can exhibit a formal solution to the last two equations in \eref{48} as 
\beq
\label{49}
C=*\,{1\over U^2}\,Ui_U \,\left({1\over U^2}\,Ui_U+{1 \over \qua}\,\delta d
\right)^{-1}
{1\over \qua}\,d* J,
\eeq
where the inverse operator in the brackets sends 
two--forms into two--forms and has to be defined according to the boundary
condition \eref{bound}. For the particular case $U^\mu(x)=N^\mu$ 
\eref{49} can be seen to reduce to \eref{24}. 

\subsection{Field theory action}

We present now an action which gives rise to the
set of equations of motion given above, and which will be the basis for the 
quantum field theory.
Its basic building block will be the PST--action,
but the two--forms $C$ are now treated as independent variables. The last
two equations in \eref{48} can be entailed by a convenient set of
lagrange multiplier fields. The consistency of this method requires to
show that these new fields do not propagate and become algebraic functions
of the physical fields. This will be the main consistency 
check for the action we propose.

We introduce as  Lagrange multipliers a doublet of two--forms
$\Sigma^I$ and a doublet of one--forms $\Lambda^I$, which are viewed as  
$U(1)$ vector potentials. Denoting collectively
the fields as $\phi=(A,C,a,\Sigma,\Lambda,\varphi_r)$, the 
proposed action, for bosons, is given by
\beq
\label{50}
S_U[\phi]=S_0+\int\left(\Lambda\,\ve\,dC-{1\over 2}\Sigma\,Ui_U\,\ve\,C
\right) -\sum_r\int d^4x\,\bar\varphi_r\left(D^\mu(\Lambda)D_\mu(\Lambda)
+m_r^2\right)\varphi_r.
\eeq
Here $S_0[A,C,a]$ is the PST--action and the covariant derivatives
on the scalars are defined as in \eref{47}, but with the vector 
potentials $\Lambda$ replacing  $A$. It is understood that $U^\mu(x)$ 
is regarded as a fixed external field and that the boundary conditions 
\eref{bound} are imposed.

The formula \eref{50} for $S_U$ can be regarded, in a sense, as the key 
equation of the paper.

The action \eref{50} possesses, in addition to the PST--symmetries for 
$a$ and $A$, the following invariances
\bea
\label{51}
&&\cases 
{
\varphi_r'=e^{-ie_r\,\ve\,\lambda}\,\varphi_r &\cr
\Lambda'=\Lambda +d\lambda, &\cr
}
\\
&&\nonumber\\
\label{52}
&&\cases
{
C'=C+dH&\cr
\Sigma'=\Sigma-dH & for\quad  $i_UdH=0,$\cr
A'=A-H&\cr
}
\\
&&\nonumber\\
\label{53}
&&\phantom{C} \Sigma'=\Sigma+Ui_U\Delta,
\eea
where $\Delta$ is a two--form doublet.
The transformations \eref{51} are ordinary $U(1)$--invariances for the 
vector fields $\Lambda$; the transformations in \eref{52} appear as
a kind of constrained string changes and are quasi--local, in the sense
that the parameter doublet $H$ is constrained by $i_UdH=0$. The last 
transformations imply that the fields $\Sigma$
are defined modulo forms parallel to $U$, as is obvious from the manner 
they appear in the action.

The equations of motion for $a$ and $A$ amount, as in the pure PST--case,
to 
\beq\label{53a}
dK=dC,
\eeq
where $K^I$ are the off--shell electromagnetic tensors defined previously.
Variation of $S_U$ with respect to $\Sigma$ and $\Lambda$ gives 
respectively
\beq
\label{54}
i_UC=0,\quad dC=J(\varphi,\Lambda),
\eeq
where $J(\varphi,\Lambda)$ are the matter current three--forms defined above 
but with $A$ replaced by $\Lambda$. The equations of motion for $\varphi_r$
are the Klein--Gordon equations, again with $A\rightarrow\Lambda$.
The equation of motion for $C$ becomes 
\beq\label{55}
K-C-d\Lambda=-{1\over 2}\left(C+*Ui_U*\Sigma\right),
\eeq
and it should determine the auxiliary fields $\Sigma$ and $\Lambda$. To show 
that this, actually, happens we make the natural gauge choice for 
\eref{53} $i_U\Sigma=0$, which is equivalent to $\Sigma =-*Ui_U*\Sigma$.
Then, defining $G={1\over 2}(C-\Sigma)$ \eref{53a}, \eref{54} and \eref{55}
imply that
$$
dG=0=i_UG.
$$
But since under \eref{52} $G$ transforms as $G\rightarrow G+dH$, with 
$i_UdH=0$, these equations mean precisely that $G$ is ``pure gauge" under
this symmetry and one can set $G=0$. The left hand side of \eref{55} must
then also vanish, and if one fixes the PST--symmetries in 
the standard way such as to give $K=F=dA+C$, one determines finally the 
Lagrange multipliers as
\bea
\label{56}
\Sigma &=&C\\
\label{57}
\Lambda &=&A,
\eea
apart from a $U(1)$ transformation \eref{51}. 

Since now $\Lambda=A$, the 
equations of motion for all the other fields amount indeed  to the set
\eref{48}, q.e.d. In particular, the physical propagating
fields are made out only of one photon and of the matter fields. 
But, since the identification \eref{57} occurs only on--shell, for a
correct identification of the observables e.g. Wilson loops, the 
electromagnetic field tensor etc. some caution is required in picking
up one vector potential or the other. As we shall see in the next section
symmetry reasons will always lead one to a unique choice.

\section{Quantum field theory}

Our formulation of a quantum field theory for dyons interacting with an 
electromagnetic field will be based on the classical action $S_U$, 
with the boundary condition \eref{bound}, in the framework of a
functional integral approach. Since the correlation functions of the 
gauge--invariant operators encode all physical information, it suffices 
to construct only these correlation
functions. The classical action depends on $U$ and, as a consequence, 
also these correlation
functions depend a priori on this external field. The main goal of this
section is to show that, if \eref{1} holds, the $U$--dependence disappears
in the correlation functions of physical fields. Manifest $SO(2)$-- and 
Lorentz--invariance then follows automatically. 

To prove $U$--independence we use a standard technique:  
path integral representations in terms of classical
particle trajectories \cite{Feynman} of determinants and Green functions. 
For definiteness we consider bosonic matter fields and comment eventually
about the fermionic case which presents, however, no additional 
difficulties. For more details on path integral representations for particles
with arbitrary spin we refer the reader to \cite{path}.

\subsection{Path integral representations}

We begin by writing the representation  for the determinant associated
to the scalar field $\varphi_r$ in \eref{50}
\beq
\label{58}
\int \{ {\cal D}\varphi_r{\cal D}\bar\varphi_r\}\,
  e^{-i\int\bar\varphi_r\left(D^2_r+m_r^2\right)\varphi_r}
={\rm det}^{-1}\left(-i\left(D^2_r+m_r^2\right)\right)
={\rm exp}\left(-tr\,\,\ln\left(-i\left(D^2_r+m_r^2\right)
\right)\right),
\eeq
where the covariant derivative on $\varphi_r$ is 
$D_{r\mu} =\left(\pa_\mu+i\,e^I_r\,\ve^{IJ}\Lambda^J_\mu\right)$. 
Apart from a normalization, the basic
path representation is given formally by 
(see e.g. \cite{GJ} for a rigorous definition):
\beq
\label{59}
tr\,\,\ln\left(-i\left(D^2_r+m_r^2\right)\right)
=-\int^\infty_0{dT\over T} \int\{{\cal D}y_r\}
\,e^{-i\int_0^T \left({1\over 4}\left({dy_r\over ds}\right)^2+m_r^2+e_r\,\ve\,
\Lambda_\mu {dy_r^\mu\over ds}\right)ds}, 
\eeq
where the integration is over all closed path, $y_r(T)=y_r(0)$. We rewrite
\eref{59} in a  more concise  form as 
$$
tr\,\,\ln\left(-i\left(D^2_r+m_r^2\right)\right)
=-\int\{{\cal D}\gamma_r\}\,e^{-i\int \Lambda\,\ve\,e_r \widetilde J_r},
$$
where the integrand in the exponent is written as a four--form and the 
{\it classical} current three--form $\widetilde J^r$ is defined as 
the PD--dual of the closed curve $\gamma_r$, parametrized by $y_r(s)$.
We absorbed in the path integral measure $\{{\cal D}\gamma_r\}$ the 
exponential prefactors in \eref{59} and included formally also the integral 
over $T$. Expanding eventually the exponential in \eref{58} we arrive to 
the identity
\bea
{\rm det}^{-1}\left(-i\left(D^2_r+m_r^2\right)\right)&=&
\sum_{l=0}^\infty{1\over l!}\left(\prod_{k=1}^l
\int\{{\cal D}\gamma_r^{(k)}\}\right) e^{-i\int \Lambda\,\ve\,e_r 
     \sum_{k=1}^l \widetilde J_r^{(k)}}\\
&\equiv& \int \{{\cal D}\Gamma_r\}\, 
e^{-i\int \Lambda\,\ve\,e_r J_r},
\eea
where the integer three--form $J_r$ represents now a sum of $l$ classical 
point particle currents, each one being of the form \eref{18a}, and the measure 
$\{{\cal D}\Gamma_r\}$ includes now also a sum over the number of loops.   

The functional integration over all $N$ matter fields leads then to
\beq
\label{60}
\prod_{r=1}^N {\rm det}^{-1}\left(-i\left(D^2_r+m_r^2\right)\right)=
\left(\prod_{r=1}^N  \{{\cal D}\Gamma_r\}\right)
e^{-i\int \Lambda\,\ve\, \sum_{r=1}^N e_r J_r}
\equiv \int\{{\cal D}\Gamma\}\, e^{-i\int \Lambda\,\ve\,J},
\eeq
where the three--forms $J^I$ correspond now to a {\it weighted sum of 
closed integer forms}, as in \eref{18} and \eref{18a}.

An analogous representation can be given for the matter two--point function
in an external gauge field $\Lambda$:
\bea
\nonumber
\langle T\bar\varphi_r(x)\varphi_r(y)\rangle_\Lambda
&\equiv&
\int \{ {\cal D}\varphi_r{\cal D}\bar\varphi_r\}\,
  e^{-i\int\bar\varphi_r\left(D^2_r+m_r^2\right)\varphi_r}\,
\bar\varphi_r(x)\,\varphi_r(y)\\
\nonumber
&=&{\rm det}^{-1}\left(-i\left(D^2_r+m_r^2\right)\right)\cdot
\left(-i\left(D^2_r+m_r^2\right)\right)^{-1}(x,y)\\
\nonumber
&=&{\rm det}^{-1}\left(-i\left(D^2_r+m_r^2\right)\right)
\int^\infty_0 dT\int\{{\cal D}y_r\}
e^{-i\int_0^T \left({1\over 4}\left({dy_r\over ds}\right)^2+m_r^2
+e_r\,\ve\,
\Lambda_\mu {dy_r^\mu\over ds}\right)ds}\\
&\equiv&
{\rm det}^{-1}\left(-i\left(D^2_r+m_r^2\right)\right)\cdot
\int\{{\cal D}\gamma_r^{(x,y)}\}\,e^{-i\int \Lambda\,\ve\,e_r  J_r^{(x,y)}}.
\label{61}
\eea
Here the path integration is over open curves $y_r(s)$ with endpoints 
$y_r(0)=x$, $y_r(T)=y$, and the integer three--forms $J_r^{(x,y)}$
are PD--dual to those curves.

\subsection{Correlators of observables and $U$--independence} 

We begin by establishing the $U$--independence of the partition function.
This is the simplest case and it encodes already the two basic ingredients
which allow to establish it for all observable correlation functions.
The first ingredient is the technical advice of path integral 
representations for 
the matter fields and the second is represented by the vanishing of the
Dirac--anomaly for {\it classical} point particles, a result which we
have already established.

\subsubsection{Partition function}

The partition function is represented by
$$
Z_U=\int \{{\cal D}\phi\}\,e^{iS_U[\phi]},
$$
where the boundary condition \eref{bound} is imposed on the functional 
integration over $C$, and appropriate gauge fixings for the local 
symmetries are understood. 

The integration over $A$ and $a$ has already been performed explicitly
in section two and gives rise to what we called the effective 
current--current action $\Gamma_0[C]$. 
The integration over $\Sigma$ gives the 
$\delta$--function $\delta(i_UC)$ while for the integration over the matter
fields we can use the path--integral representations given above. Putting 
everything together one obtains
\bea
\nonumber
Z_U&=&\int\{{\cal D}\Gamma\}\{ {\cal D}C\}\{{\cal D}\Lambda\}\,\delta(i_UC)
\,e^{i\Gamma_0[C]+i\int\Lambda\,\ve\,\left(dC-J\right)}\\ 
\nonumber
&=& \int\{{\cal D}\Gamma\}\{ {\cal D}C \}\,\delta\left(i_UC\right)\,
\delta\left(dC-J\right)\,e^{i\Gamma_0[C]}. 
\eea
The $\delta$--functions restrict the variable 
$C$ to $dC=J,\,i_UC=0$ (remember that $J$ satisfies $dJ=0$ identically)
which, taken together with the boundary condition \eref{bound}, admit 
a unique solution for $C$; since $J$ is now a classical current this 
solution is 
given by \eref{25b} and \eref{26} and we call it $C^I(J,U)$. Therefore
\beq
Z_U=\int\{{\cal D}\Gamma \}\,e^{i\Gamma_0[C(J,U)]},
\label{61a}
\eeq	
which is our final expression.

An arbitrary change in the external field from $U\rightarrow U'$ corresponds 
now, for fixed $J$, just to a string--change $C(J,U')=C(J,U)+dH_{U,U'}$,
where the $H_{U,U'}$ are weighted integer one--forms which correspond to 
the three--volumes bounded  by the surfaces associated respectively
to $U$ and $U'$ according to \eref{25b}. But under a string--change
$\Gamma_0$ changes by the Dirac--anomaly   
$$
Z_{U'}=\int\{{\cal D}\Gamma \}\,e^{i\left(\Gamma_0[C(J,U)]
+{1\over 2}\int J\,\ve\,H_{U,U'}\right)},
$$ 
and if \eref{1} holds $exp\left({i\over 2}\int J\,\ve\,H_{U,U'}\right)=1$. 
Thus $Z_U=Z_{U'}$.

As long as one considers observable correlation functions, i.e. correlation
functions of invariant operators the strategy presented here will always
lead to the desired result. Below we work out the most significant
cases.

\subsubsection{Current correlation functions}

In this subsection we consider the correlation functions of the 
individual matter currents ($r=1,\ldots,N$),
\beq\label{62}
j_{r\mu}=i\bar\varphi_r D_\mu(\Lambda)\varphi_r+c.c.
\eeq
In writing this expression we solved already the ambiguity regarding
the choice between $A$ and $\Lambda$. Here one has to choose $\Lambda$
because the corresponding expressions with $A$ would violate the 
PST--symmetries. 

The current correlators are conveniently encoded
by a generating functional which can be constructed in a standard
way. One introduces $N$ external currents $L_r$, one--forms, and
performs in the Klein--Gordon term in $S_U$ the replacement
\beq
\label{63}
e_r\,\ve\,\Lambda \rightarrow  e_r\,\ve\,\Lambda +L_r,
\eeq
without affecting the term $\int\Lambda\,\ve\,dC$ and the other terms  of the 
action. We call the action obtained in this way $S_U[\phi,L]$. 
The terms in $S_U[\phi,L]$ which are
linear in $L_r$ multiply simply the currents \eref{62}, while the terms
which are quadratic in $L_r$ -- contact terms -- are needed to ensure
the global invariances $[U(1)]^N$, i.e. the individual current conservations.

The correlation functions are now obtained, via  differentiation w.r.t.
the $L_r$, from the generating functional
$$
e^{iW[L]}=\int\{{\cal D}\phi\}\,e^{iS_U[\phi,L]}.
$$
This functional integral can be analyzed in the same manner as the 
partition function. The Klein--Gordon determinants can be
obtained from \eref{60} with the substitution \eref{63} and their
product can be written as 
$$
\int\{{\cal D}\Gamma\}\, e^{-i\int\left( \Lambda\,\ve\,J -\sum_r L_rJ_r
\right)}.
$$
The integration over the other fields can be handled as above and one obtains
\bea
e^{iW[L]}=
\nonumber
&=&\int\{{\cal D}\Gamma\}\{ {\cal D}C\}\{{\cal D}\Lambda\}\,\delta(i_UC)
\,e^{i\left(\Gamma_0[C]+\int\Lambda\,\ve\,\left(dC-J\right)
+\sum_r \int L_rJ_r\right)}\\ 
\nonumber
&=& \int\{{\cal D}\Gamma\}\{ {\cal D}C \}\,\delta\left(i_UC\right)\,
\delta\left(dC-J\right)\,e^{i\left(\Gamma_0[C]+
\sum_r \int L_rJ_r\right)} \\ 
\nonumber
&=&\int\{{\cal D}\Gamma \}\,e^{i\left(\Gamma_0[C(J,U)]
+\sum_r \int L_rJ_r\right)}. 
\eea
Since the integrals $\int L_rJ_r$ are $U$--independent the 
Dirac--anomaly is the same as in the case of the partition function. 
The $U$--independence of $W[L]$ then also follows.  

\subsubsection{Electromagnetic field correlation functions}

We noticed already that the off--shell electromagnetic tensor
is represented by the two--form doublet $K^I$ which is given as a function
of $A$, $C$ and $a$ in \eref{13}. It satisfies in particular
the pseudo self--duality relation
\beq
\label{64}
K=*\,\ve\,K,
\eeq
which is necessary to avoid a doubling of the electromagnetic degrees of 
freedom. The alternative for $K$, in terms of $\Lambda$, would in this
case be given by $d\Lambda+C$ because on--shell we have $K=dA+C$ and 
$A=\Lambda$. But the expression $d\Lambda+C$ violates the quasi--local
invariances \eref{52} and, moreover, it does not satisfy \eref{64}.

Therefore, if we introduce as external currents a doublet of two--forms, $L^I$,
the generating functional for the electromagnetic field correlators
could be based on the action
\beq
\label{65}
S_U+ \int K\,\ve\,L,
\eeq
where the $\ve$--tensor has been introduced only for notational 
convenience. The resulting generating functional would then 
automatically satisfy the pseudo self--duality invariance
\beq
\label{65a}
W[-*\ve\,L]=W[L],
\eeq
i.e. generate correlation functions only for a single field strength
tensor.

The expression \eref{65} represents already the correct starting
point, apart from contact terms. As in the case
of current correlators these terms, quadratic in
$L$, have to be fixed in order to respect the symmetries of the underlying 
theory. 
The problem related with \eref{65} is that the $K$'s are invariant under the 
PST--symmetry \eref{9} only on--shell. 
Therefore, the direct use of \eref{65} would 
introduce in the $K$--correlators a spurious dependence on the 
unphysical auxiliary field $a$. 

The problem can be solved by basing the generating functional
on the following action 
\bea
\nonumber
S_U[\phi,L]&=&S_U +\int\left(K\,\ve\,L+{1\over 2}L\,{\cal P}(v)\,L
                   -{1\over 4}L*L\right)\\
\nonumber
&\equiv& S_L+ \int\left(\Lambda\,\ve\,dC-{1\over 2}\Sigma\,Ui_U\,\ve\,C
\right) -\sum_r\int d^4x\,\bar\varphi_r\left(D^\mu(\Lambda)D_\mu(\Lambda)
+m_r^2\right)\varphi_r,
\eea
where $S_L$ represents a deformation of the PST--action 
$$
S_L=S_0[A,C+L,a]+{1\over 2}\int\left( C\,\ve\,L-{1\over2}L*L\right).
$$
The quadratic terms we have added depend only on $L$ and $a$, therefore 
they influence only the PST--symmetry \eref{9}, which sends $a\rightarrow
a+\varphi$. From the form of $S_L$ one sees that it is invariant under
the transformations \eref{9} if one replaces in the  $A$--transformations
$C$ with $C+L$. This ensures again the decoupling of the auxiliary field
$a$. To achieve this decoupling only the $a$--dependent term 
${1\over 2}\int L\,{\cal P}(v)\,L$ would be required, but this term alone 
is not invariant under pseudo--duality, i.e. under 
$$
L\rightarrow -*\ve\,L.
$$
It is only the combination ${1\over 2}\int\left( 
L\,{\cal P}(v)\,L-{1\over2}L*L\right)$ which is invariant.

The generating functional for the field strength correlators, defined as
$$
e^{iW[L]}=\int\{{\cal D}\phi\}\,e^{iS_U[\phi,L]}
$$
with $S_U[\phi,L]$ specified above, entails, therefore, all the required
symmetry properties, in particular \eref{65a}. 
It remains to show that it is $U$--independent. 
To this order we perform first the functional integration over the fields 
$A$ and $a$, which appear only in $S_L$,
$$
e^{i\Gamma_L[C]}\equiv\int \{ {\cal D}A\}\{{\cal D} a\}\,e^{iS_L}.
$$
The result for $\Gamma_L[C]$ can be easily read from the analogous 
functional integration over the PST--action, which gave the effective 
action $\Gamma_0[C]$,
$$
\Gamma_L[C]=\Gamma_0[C+L] +{1\over2}\int\left(C\,\ve\,L-{1\over2}L*L
\right).
$$
What matters eventually is how this expression transforms under a generic
string change $C\rightarrow C+dH$. By direct inspection one finds that it
transforms precisely as $\Gamma_0[C]$,
\beq
\label{66}
\Gamma_L[C] \rightarrow \Gamma_L[C] +A_D[H],
\eeq
where $A_D[H]$ is the usual Dirac--anomaly. This is a consequence of the 
fact that also $S_L$ carries the same Dirac--anomaly as $S_0$.

The integration over the remaining fields is carried out as in the
case of the partition function and the result for the generating functional
is obtained from the
expression for $Z_U$ through the replacement $\Gamma_0\rightarrow 
\Gamma_L$,
$$
e^{iW[L]}=\int\{{\cal D}\Gamma \}\,e^{i\Gamma_L[C(J,U)]}.
$$
$U$--independence then follows from \eref{66}.

\subsubsection{Wilson loops}

Wilson loops are associated to a set of $N$ closed space--time curves 
$\beta_r$, one for each dyon, and represent the interaction between the 
vector potential and the particles circulating in the loops. 
The correlator of the corresponding observables is defined as
\beq
\label{67}
{\cal W}[\beta]= \left\langle T\prod_{r=1}^N e^{-ie_r\,\ve\,\oint_{\beta_r}
\Lambda}\right\rangle=\left\langle T \, e^{-i \int \Lambda\,\ve\,J(\beta)}
\right\rangle.
\eeq
Here we defined the closed three--form doublet $J^I=\sum_r e_r^I J_r(\beta)$, 
where the three--forms $J_r(\beta)$ are the PD--duals of the loops $\beta_r$.

In the definition of the correlators we have to use again  $\Lambda$ 
instead of $A$ since Wilson loops constructed with the latter would violate
the PST--symmetries. On the other hand, since the loops $\beta_r$ are 
closed, the classical $U(1)$--invariances of $\Lambda$ would allow 
also for arbitrary real weights of the $N$ individual terms in the
exponentials in \eref{67}. But the corresponding 
quantum correlators would be string--dependent, that is $U$--dependent,
as one expects also on the basis of Dirac's original argument for 
charge quantization. Thus these weights have to take on {\it integer} 
values.

The functional integral representation for \eref{67} is
$$
{\cal W}[\beta]=\int\{{\cal D}\phi\}\,e^{iS_U}
\,e^{-i \int \Lambda\,\ve\,J(\beta)}.
$$
Once one represents the matter determinants as in \eref{60}, one sees that 
the presence of the Wilson loops amounts merely to a replacement of
the  closed currents $J$ with the closed currents $J+J(\beta)$. Thus
$$ 
{\cal W}[\beta]=\int\{{\cal D}\Gamma \}\,e^{i\Gamma_0[C(J+J(\beta),U)]}.
$$
But since $J+J(\beta)$ is also a weighted integer current the
Dirac--anomaly is again a multiple of $2\pi$ and the correlators are
$U$--independent. 

\subsubsection{Mandelstam--string observables}

We consider here correlation functions corresponding to a particular 
type of neutral composite fields, formed by a certain number  of particles and
antiparticles, with support on a bounded string. 
Due to the global $[U(1)]^N$ invariances only correlation functions
in which each particle appears together with its own antiparticle
are non--vanishing. 

Here we work out only the one--point Green function for a ``meson" field, 
involving say the $r$-th species, generalizations to multiple states being 
straightforward. The Green function we consider is specified by
the positions $x$ and $y$ of the two particles and by an open
curve $\beta_{x,y}$, the ``Mandelstam--string" \cite{Mandel}, 
which connects the two points. 
The corresponding correlator is given by
\bea\nonumber
G_r[\beta_{x,y}]&=&\left\langle T\,\bar\varphi_r(x)\,e^{ie_r\,\ve\,
\int_{\beta_{x,y}}
\Lambda}\,\varphi_r(y)\right\rangle=
\left\langle T\,\bar\varphi_r(x)\,e^{i\int \Lambda \,\ve\,e_r J_r
(\beta_{x,y})}\,\varphi_r(y)\right\rangle\\
&=&
\nonumber
\int\{{\cal D}\phi\}\,e^{i\left(S_U +\int \Lambda \,\ve\,e_r J_r
(\beta_{x,y})\right)}\, \bar\varphi_r(x) \,\varphi_r(y).
\eea
The insertion of the Mandelstam line--integral along $\beta_{x,y}$,
with vector potentials $\Lambda$, is needed to realize the local 
electromagnetic invariances \eref{51}. 

The evaluation of the correlator can be based on the expansion 
\eref{61}. The determinant in that formula combines with the determinants
of the other $N-1$ charged fields and leads again to \eref{60}. One obtains
\bea
\nonumber
G_r[\beta_{x,y}]&=&
\int\{{\cal D}\gamma_r^{(x,y)}\}
\int\{{\cal D}\Gamma\}\{ {\cal D}C\}\{{\cal D}\Lambda\}\delta(i_UC)
e^{i\Gamma_0[C]+i\int\Lambda \ve\left(dC-\left(J+
e_r\left(J_r^{(x,y)}-J_r(\beta_{x,y})\right)\right)\right)}\\
\label{68}
\phantom{x}&&\phantom{y}
\eea
Since $J_r^{(x,y)}$ and $J_r(\beta_{x,y})$ are PD--dual to the open curves
$\gamma_r^{(x,y)}$ and $\beta_{x,y}$ respectively, the total currents
$J+e_r\left(J_r^{(x,y)}-J_r(\beta_{x,y})\right)$ are again
closed and weighted integer forms. 
This is sufficient to guarantee $U$--independence once more.

Generalizations to $n$--point Green functions  are straightforward since the 
second row in formula \eref{61} admits immediate extensions to 
arbitrary polynomials in  $\{\varphi_r\}$, in terms of multiple contractions.

\subsubsection{Charged fields}

We wish here to outline the construction of Green functions for charged
fields, following the strategy developed in \cite{PAM} (for all technical 
details we refer the reader to this reference). The fields we consider are 
required to be invariant under local $U(1)$ gauge transformations 
which vanish at infinity, 
but to transform non--trivially, i.e. to be charged, with respect to the 
global $U(1)$ transformations. In agreement with Strocchi's theorem 
\cite{St}, these requirements imply that the fields are non--local, 
more precisely that their support 
is non--compact and extends  to infinity.

There are two natural candidates for such fields. The first is obtained 
multiplying (``dressing") a charged matter field $\varphi_r (x)$ by a 
Mandelstam--string phase factor 
$$
exp\left(i \int_{\gamma_x} e_r\,\ve\,
\Lambda\right).
$$ 
$\gamma_x$ is here a curve starting from $x$ and reaching 
infinity along some fixed space--direction at fixed time,
where it meets a ``compensating charge".

The second candidate field is obtained, according to Dirac's original 
proposal \cite{PAMdirac}, dressing $\varphi_r (x)$ by a Coulomb phase factor
$$
exp\left(i e_r\, \ve \int d^4y\, \Lambda_\mu(y) E^\mu_x(y)\right).
$$
$E_x$ is a classical Coulomb 
electric field, generated by a unit charge located at $x$, 
with support in a fixed--time three--space and satisfying 
$\pa_\mu E^\mu_x(y)=\delta^4(x-y)$ \footnote{In lattice abelian gauge 
theories without dynamical monopoles, 
it has been shown rigorously \cite{FP} that charged states 
and charged quantum fields can be reconstructed out of euclidean Green 
functions, derived as expectation values of globally neutral products of 
charged fields defined according to Dirac's proposal.}. 
More precisely
\bea
\vec E_x(y)&=& {1\over 4\pi}\,\delta (y^0)\,{\vec y -\vec x\over
|\vec y -\vec x|^3}\\
E_x^0(y)&=&0.
\eea 
From a physical point of view 
Dirac's recipe describes a charged particle dressed with its Coulomb 
photon cloud.  The latter appears energetically more favorable then the 
Mandelstam--string proposal in Q.E.D.--like theories, as we know that in such 
theories the electromagnetic field associated classically to a charged 
particle is indeed Coulomb--like.

However, neither of these candidates appears to be suitable in a theory 
where electric and magnetic dynamical charges coexist.
The problem arising with Mandelstam--strings is related with 
infrared divergences (they are unbounded) and it can be understood as follows. 
Consider e.g. the ansatz two--point function for a charged particle created 
at $x$ and annihilated at $y$
$$
\left\langle \bar\varphi_r(x) e^{i\,e_r\, \ve \int_{\gamma_x}
\Lambda} \varphi_r(y) e^{-i\, e_r\, \ve \int_{\gamma_y} \Lambda} \right\rangle.
$$
It involves two strings of infinite length, one starting from $x$ and the 
other from $y$. Loosely speaking 
the strings would have infinite positive self--energies and infinite 
negative interaction--energy as a consequence of the interaction mediated 
by the photons. More concretely, the effective action 
$\Gamma_0[C(J,U)]$ diverges if the currents $J$ correspond to unbounded
curves, like the above Mandelstam--strings.
The self--energies could be in principle removed by a 
suitable multiplicative point--independent renormalization. But the 
interaction energy can not be removed as it depends on the distance 
between $x$ and $y$, and thus the ansatz Green function would be infrared 
divergent. (More precisely, in the euclidean formulation renormalizing away 
the interaction energy would spoil Osterwalder--Schrader positivity of the 
correlation function, a property which is crucial for the reconstruction
of the quantum field, see e.g. \cite{GJ}.)

On the other hand, Dirac's construction can not be used in a theory of 
dyons  
because the field $E_x$ would give rise to a smooth non integer current 
$J$ and the Dirac--anomaly would no longer be an integer multiple of
$2\pi$. This would then imply an unphysical dependence of the Green 
functions on Dirac--strings.

The recipe presented in \cite{PAM}, which overcomes these difficulties, 
consists in replacing a 
single Mandelstam--string $exp(i e_r\, \ve\int_{\gamma_x}\Lambda )$ 
by a sum over Mandelstam--strings weighted by a suitable measure 
$d\mu\{\gamma_x\}$. This measure is 
supported on strings $\gamma_x$ which fluctuate so strongly that with  
probability one the corresponding interaction--energy is finite, and       
it entails the additional feature that on average
it reproduces just the Coulomb field $E_x$. Therefore, although on small 
scales the ``photon cloud" associated to the charged field 
$$
\varphi_r(x) \int d \mu\{\gamma_x\}\, e^{i e_r\, \ve \int_{\gamma_x} \Lambda}
$$
respects the Dirac quantization condition, at large 
distances on average it scales to the classical Coulomb field $E_x$.

The Green functions for charged fields are eventually given by
$$
\left\langle T\prod_i \varphi_{r_i}(x_i) \int d \mu \{\gamma_{x_i}\}\,
e^{i \,e_{r_i} \ve \int_{\gamma_{x_i}} \Lambda} \right\rangle,
$$
where $\varphi$ denotes $\varphi$ or $\bar\varphi$ and total charge 
neutrality is understood. 

The strategy developed above can now be used in
the same fashion to prove Dirac--string independence of these correlation 
functions too.

\subsection{Comments}

The procedure adopted above to prove the string--independence of the 
invariant correlation functions  works equally well for fermions.
The corresponding classical action is obtained from \eref{50}
upon replacing the Klein--Gordon action with the Dirac action, and
the relevant path--integral representations are 
available also for fermions. For example, the fermionic 
determinants admit expansions of precisely the same type as \eref{60}, 
what changes is only the functional  measure over the paths \cite{ZW2}. 
Appropriate expansions for the fermionic propagators in an external 
gauge field are also available \cite{path}.

Our formulation for the quantum field theory of dyons, based on the action
\eref{50}, furnishes string--independent correlators for all gauge--invariant 
operators. It may be worthwhile to notice that correlators of operators 
which are not 
invariant under {\it local} gauge transformations, instead, turn out to 
vanish. 
Consider, for example, the scalar two--point function $\langle T
\bar\varphi_r(x)\varphi_r(y)\rangle$. This is obtained from \eref{68}
upon dropping the current $J_r(\beta_{x,y})$. The integration over 
$\Lambda$ leads then to the $\delta$--function $\delta\left(dC-\left(J+e_r
J_r(\beta_{x,y})\right)\right)$ which has no solution since the current is 
now no longer closed. This would then give a vanishing 
correlator, in agreement with Elitzur's \cite{El} theorem\footnote
{The same happens formally in ordinary Q.E.D. if one does
not fix the $U(1)$--symmetry.}. 
The point is that in the case of currents which are not closed, 
occurring in the presence of non invariant operators, it is 
necessary to fix the local U(1)--symmetries \eref{51}. But this leads then 
to {\it smooth} non integer currents and the correlators become 
$U$--dependent. 

A word is in order for what concerns the configurations which we called
in the classical point particle theory ``exceptional". These configurations
appear indeed also in the quantum field theory; taking, for example, the 
expansion of the partition function in \eref{61a}, they occur when some curve
$\gamma_r$ intersects the string $C_s(J_s,U)$ associated to another 
curve $\gamma_s$. In this case  $\Gamma_0[C(J,U)]$ 
would be ill--defined. However, every regularization which preserves the 
integrality of the currents gives the same result for 
$exp(i \Gamma_0[C(J,U)])$. 
More concretely, the problem can be solved by performing in the expansion 
\eref{61a} a string change from $C_s(J_s,U)$ to an arbitrary regular string
$\widetilde C_s(J_s)$ (see section three). 
Since $exp(i \Gamma_0[C(J,U)])$ is string--independent,
the regularized expression  $exp(i \Gamma_0[\widetilde C(J)])$ equals the 
original one for regular configurations and is, moreover, independent
of the regularization. This regularization procedure fails only if the
{\it curves} themselves intersect each other; but, in the models considered
here, these configurations form a set of measure zero in the path integral.

A classical field theory action of the kind \eref{50}
can be given also for the $Z_4$ theory. In this case one has only one 
vector potential $A^2$, and only one lagrange multiplier field, 
$\Sigma^1=C^1$, is needed to impose $dC^2=J^2$. 
The fields are now
$\phi=(A^2,\Lambda^1,C^I,\varphi_r)$ and the counterpart of \eref{50} is
$$
S_U^{Z_4}[\phi]=\int {1\over 2}\,F^2*F^2 +\Lambda^1dC^2-C^1Ui_UC^2
 -\sum_r\int d^4x \,\bar\varphi_r\left(D^\mu D_\mu+m_r^2\right)
\varphi_r,
$$
where $F^2=dA^2+C^2$, and the covariant derivative is here 
$D_\mu=\pa_\mu+i\left(e_r^1A^2_\mu-e_r^2\Lambda^1_\mu\right)$. For
the $C^I$ we assume the usual boundary conditions \eref{bound}. 
The symmetries of 
this action are analogous to the ones given in \eref{51}--\eref{53}
for the $SO(2)$--theory and, together with the 
equations of motion, they determine the  auxiliary fields as follows
\bea
\label{69}
d\Lambda^1+C^1&=&*F^2\\
\label{70}
dC^I&=&J^I(A^2,\Lambda^1,\varphi_r), \quad i_UC=0.
\eea
Taking the matter field equations into account and setting 
$\Lambda^1\equiv A^1$, one recovers precisely the system of equations
of motion of the $SO(2)$--theory, \eref{48}. 

The corresponding
quantum field theory can be analyzed in the same way as the $SO(2)$--theory
and it turns out to be consistent if \eref{2} holds.

It is worthwhile to notice a puzzle which arises in comparing the 
inequivalent quantum field theories based on $S_U$ and $S_U^{Z_4}$ 
respectively. As we noted in section 4.2. the difference arises 
in the effective actions 
\beq
\label{71}
\Delta \Gamma_0=\Gamma^{Z_4}_0-\Gamma_0={1\over 2}\int C^1C^2.
\eeq
Since in both quantum field theories the strings are chosen to satisfy 
$i_UC=0$ and one has the identity decomposition 
$1={1\over U^2}\left(i_UU-Ui_U\right)$, one would conclude 
$$
C^1C^2={1\over U^2}\,C^1\left(i_UU-Ui_U\right)C^2=0.
$$
The point is that in deriving \eref{71} we performed integrations by 
parts and ignored boundary terms at infinity; non vanishing boundary terms
arise, in general, if the strings extend to infinity, as does $C[J,U]$. 
Strings
extending to infinity are unavoidable if the curves themselves extend to
infinity, as in the classical point particle theory. But in the partition 
function or in the correlation functions of local observables in the 
quantum field 
theory the curves are closed and one can also choose {\it compact} surfaces as 
strings. On the other hand, since $exp(i\Gamma_0)$
is string--independent, one can switch from $C[J,U]$ to arbitrary
compact strings $C_{com}[J]$. Since now curves {\it and} strings are 
compact no boundary terms arise and one gets indeed \eref{71}. But then
$C_{com}^1C_{com}^2$ is different from zero and one concludes that the two 
quantum field 
theories coincide only if the charges satisfy the stronger relations 
\eref{stronger}, as anticipated in section 4.2. 

  The extension of our formulation to include a coupling with an external
classical gravitational field, represented by a metric $g_{\mu\nu}(x)$,
does not meet additional difficulties. Since the classical field theory
action is formally $SO(1,3)$--invariant, the metric can be introduced
in a standard fashion, via the minimal coupling, and the resulting action
is formally diffeomorphism invariant; the specification ``formally"
is referred to the presence of the external field $U$. The quantum 
correlators are defined again by the functional integral and one has only
to worry about the dependence of the correlators on $U$. With respect
to this issue the first key observation is that the path--integral 
representations for determinants and correlators, of the type
\eref{60} and \eref{61}, hold true. Since the exponentials involving 
$\Lambda$ are written as integrals of products of {\it forms}, they do not
acquire any metric dependence; it is only the path--integral measure which 
involves now also the metric. This leads then to expansions which are
of the same form as \eref{61a}, but clearly $\Gamma_0$ depends now also
on $g_{\mu\nu}$. The second key observation is that, nevertheless,
the Dirac--anomaly is metric--independent. This is due to the fact that
the concept of PD--duality is metric independent and that the 
Dirac--anomaly is written as the integral of a product of {\it forms}.
Its exponential equals, therefore, again unity and the correlators are
$U$--independent.

\section{Duality symmetries and  equivalences} 

The aim of this section is to establish which theories of a fixed
number $N$ of species of interacting dyons are physically equivalent to 
each other. 
Different actions can be connected by field redefinitions, upon elimination
of auxiliary fields or through an $S$--duality transformation of a 
$p$--form potential (or by a combination of these operations).
These operations suffice to establish the equivalences between 
the actions considered in this section. Theories based on classical 
actions which can be connected in one of these ways are  
physically equivalent also at the quantum level.

\subsection{S--duality for the vector potentials $A^I$}

We saw in section 4.3 that, by applying an $S$--duality transformation of the
vector potential $A^2$ in the $Z_4$--theory, one can obtain a theory in which 
the electric and magnetic charges are interchanged. The electric
charges are now magnetically coupled and the magnetic charges electrically.
This means that $Z_4$--theories with charges respectively  $(e_r,g_r)$
and $(g_r,-e_r)$, satisfying \eref{2}, are equivalent. 

In the $SO(2)$--theory both types of charges are electrically {\it and} 
magnetically coupled. This indicates that under an $S$--duality of 
the vector potentials $A^I$, the PST--action should turn into itself. This 
is indeed what happens. In fact, consider instead of the PST--action
\eref{pst} the action 
$$
S_H= {1\over 2}\int(H+C)\,{\cal P}(v)\,(H+C) + H \,\ve\,
(d\Lambda+C),
$$
where $H^I$ is a doublet of two--forms and $\Lambda^I$ a doublet of
one--forms, dual to $A^I$. Functional integration over $\Lambda$ leads to $dH=0
\Rightarrow H=dA$ and one is back to the PST--action. On the other hand,
one can now perform the integration over $H$  
$$
 \int \{ {\cal D} H\}\,e^{iS_H} \equiv e^{i\widehat S_0[\Lambda,C,a]}.
$$
The integral is Gaussian, and to perform it one has only to know the inverse 
of the $a$--dependent operator ${\cal P}(v)$ on two--form doublets
$$
{\cal P}^{-1}(v)=-{1\over 4}\,{\cal P}(v).
$$  
The result is indeed
$$
\widehat S_0[\Lambda,C,a]= S_0[\Lambda,C,a],
$$
i.e. the PST--action for the dual vector potentials $\Lambda$. This
generalizes the result of \cite{MPS}, for the case of non vanishing currents.

Therefore, an $S$--duality transformation on $A$, contrary to what happens
in the $Z_4$--theory, sends the PST--action into itself, and 
does not lead to a new formulation. Moreover, this statement does not
require any quantization condition on the charges. Clearly, the same result
holds also for the total classical field theory action $S_U$.

\subsection{S--duality for the scalar $a$}

A new formulation can be obtained if one performs an $S$--duality on the
scalar auxiliary field $a$. This transformation has been performed in
\cite{MPS}, and we report here only the result. Due to the complicated
dependence on $a$ of the PST--action, it seems hard to 
implement this duality transformation at the level of the functional 
integral; one has to proceed through the equations of motion.
This will lead in the dual action to the appearance of 
a {\it quasi}--local symmetry, needed to obtain the duality relation 
$F=*\ve F$. Apart from this difference, the dual action shares the
same good properties with the PST--action, i.e. manifest $SO(1,3)$-- and
$SO(2)$--invariance.  

The field dual to $a$ is a two--form potential $b={1\over2}\, dx^\mu dx^\nu
b_{\nu\mu}$, the relation being given by
$$
da=*\,db. 
$$

To construct the dual PST--action one defines the one--forms $h=*\,db$ and
$u={h\over \sqrt{-h^2}}$, such that $u^2=-1$; $u$ is the counterpart of the
one--form $v$. The dual action is $(F=dA+C)$
\beq
\label{dual}
\widetilde S_0[A,C,b] ={1\over 2}\int F\,{\cal Q}(u)\,F + 
dA \,\ve\,C,
\eeq
where ${\cal Q}(u)$ is a symmetric operator which acts in the space of 
two--forms and on the $SO(2)$--indices as
$$
{\cal Q}^{IJ}(u)=*u\,i_u\,\delta^{IJ} -\left(u\,i_u-{1\over 2}\right)\ve^{IJ}.
$$
Apart from the formal analogies between \eref{pst} and \eref{dual} one
should note, however, that the latter is {\it not} obtained from the former
with the simple substitution $v\rightarrow u$. One has rather
$$
{\cal Q}(u)+{\cal P}(u)=*.
$$
The symmetries of this action are
\bea
\label{72}\delta A&=& d\Lambda \\
\label{73}\delta b&=& dc \\
\label{74}\delta A&=& \chi,\quad {\rm with}\quad i_ud\chi=0\\
\label{75}\delta A&=& i_V i_uu\left(F-*\ve F\right),  \quad 
            \delta b= i_V db.
\eea
$c$ is a one--form, $\Lambda$ a doublet of scalars,  $V$ a
vector field, and $\chi$ a constrained doublet of one--forms which
entail now quasi--local transformations, as anticipated above. The 
symmetries \eref{73} and \eref{75} allow to fix $b$ completely, which 
becomes again a non--propagating auxiliary field. The symmetries \eref{74} 
allow to obtain the pseudo self--duality relation $F=*\ve F$. This implies
that the dual action describes the same classical dynamics as the 
PST--action.

Even if \eref{dual} has not been obtained from the PST--action via functional
integration, on can construct a quantum field theory, and a 
point particle theory, based
on the dual action, in precisely the same way as we did for the PST--action:
the resulting quantum field theories are, indeed, identical. 
The principal common 
features are that the two classical actions carry the same Dirac--anomaly, 
and entail the same effective action $\Gamma_0$, as shown in the 
appendix\footnote{In the calculation 
of $\Gamma_0$ from the dual action, which can be carried out using an 
identity similar to \eref{ident}, the symmetries 
\eref{74}, being quasi--local, do not need a gauge--fixing.}. 

The off--shell electromagnetic fields, the counterparts to \eref{13}, 
are now given by
$$
\widetilde K= F+*\ve\,u\,i_u\left(F-*\ve\,F\right),
$$
and have the same properties as the two--forms $K$, but now referred to
the symmetries \eref{72}--\eref{75}. 

Under an $S$--duality of the vector potentials $A$, also the dual 
PST--action goes into itself, in particular one has ${\cal Q}^{-1}(u)
=-{1\over 4}\,{\cal Q}(u)$.

The advantages of the original PST--action w.r.t. the dual action 
are represented essentially by its simpler symmetry structure, in 
particular by the absence of quasi--local symmetries. 
The most significant difference between the two actions relies in the
formulae \eref{10} and \eref{74}. The symmetry \eref{10} eliminates one
transverse degree of freedom from each $A^I$, such that each $A^I$ carries
one transverse physical photon degree of freedom. On the other hand, the 
quasi--local symmetry \eref{74} for the dual action does not eliminate 
degrees of freedom; it allows, rather, to choose boundary conditions for the 
equations of motion, which reduce the second--order equations for the vector
potentials, to the pseudo self--duality relation $F=*\ve\,F$.

As shown below, the dual action plays a significant role in 
establishing the relation 
between formulations with a single vector
potential and formulations with a doublet of vector potentials.

\subsection{Equivalence between the Zwanziger-- and PST--actions}

The classical field theory action introduced by Zwanziger \cite{ZW1}
depends only on a doublet of vector potential one--forms $\Lambda^I$,   
and on the matter fields $\varphi_r$. It is local and manifestly 
$SO(2)$--invariant, but breaks explicitly Lorentz--invariance, due to the 
appearance of a constant vector $n_\mu$, $n^2=-1$. It reads
\beq
S_{ZW}[\Lambda,\varphi]= 
{1\over 2}\int\,  d\Lambda\,{\cal Q}(n)\,d\Lambda 
-\sum_r\int d^4x\,\bar\varphi_r\left(D^\mu(\Lambda)D_\mu(\Lambda)
+m_r^2\right)\varphi_r,
\eeq
where the contribution of the matter fields is written as in \eref{50}.
The operator ${\cal Q}(n)$ is the one appearing in the {\it dual}
PST--action, but evaluated at $u^\mu=n^\mu$. 

The equivalence between this action and the classical original PST field 
theory action $S_U$ \eref{50} can be understood as follows. Since in the
PST--approach the fields $a(x)$ and $U(x)$ can be chosen arbitrarily,
the first because it is an auxiliary field and the second because the
quantum theory is $U$--independent, we choose
\bea
\nonumber
a(x)&=&n^\mu x_\mu \quad \Rightarrow v^\mu=n^\mu\\
\nonumber
U^\mu(x)&=&n^\mu.
\eea
The action $S_U$ becomes then a functional of $A,\Lambda,C,\Sigma,$ and
$\varphi_r$, which we call $S_n$.
Since the fields $\Lambda$ and $\varphi_r$  have the same symmetry
properties, i.e. $U(1)$--invariances, in $S_{ZW}$ and $S_n$,  it is natural
to try a functional integration over $C,A$ and $\Sigma$ of $S_n$:
$$
e^{iS[\Lambda,\varphi]}\equiv
\int\{{\cal D} C\}\{ {\cal D}A\}\{{\cal D}\Sigma\}\,e^{i
S_n[A,\Lambda,C,\Sigma,\varphi]}. 
$$
It is understood that the integrations have to be carried out
performing appropriate gauge fixings. 

Since in the action $S_U$ we have now to set $U=v=n$, due to the presence
of the term $-{1\over 2}\int\Sigma\,ni_n\,\ve\,C$ we can eliminate from
the PST--action all terms proportional to $i_nC$ (shifting $\Sigma$). 
This leads to  
\bea
S_n&=&\int {1\over 2}\,dA\,{\cal P}(n)\,dA+ {1\over 2}\,C*C
+\left(*\ve\,dA-d\Lambda-{1\over 2}\Sigma\,ni_n\right)\ve\,C\\     
&\phantom{=}&-\sum_r\int d^4x\,\bar\varphi_r\left(D^\mu(\Lambda)D_\mu(\Lambda)
+m_r^2\right)\varphi_r.
\eea
We perform first the gaussian integration over $C$, which leads to 
an action  which is quadratic in $\Sigma$. The integration over $\Sigma$
is again gaussian, and to perform it one has only to take the 
gauge--fixing $i_n\Sigma=0$ into account. After these integrations one 
obtains as intermediate result 
$$
\int\{{\cal D} C\} \{{\cal D}\Sigma\}\,e^{iS_n}  
=e^{i\left(S_{ZW}[\Lambda,\varphi]
+{1\over 2}\int(dA-d\Lambda)\ve\,ni_n (dA-d\Lambda)\right)}.
$$
Since there exist linear gauge--fixing conditions for the symmetries
\eref{8} and \eref{10}, one can perform the shift $A\rightarrow A+\Lambda$,
and the integral over $A$ gives a constant. Therefore, the result is
$$
S[\Lambda,\varphi]=S_{ZW}[\Lambda,\varphi],
$$
which shows the equivalence between the Zwanziger action and the
original PST--action. Notice, in particular, that the operator 
${\cal P}(n)$ of the original PST--action has turned in the
operator ${\cal Q}(n)$.

Similarly one can ask if there exists an analogous relation between
the Zwanziger action and 
the dual PST field theory action $\widetilde S_U$. The latter
is obtained from $S_U$ in \eref{50} upon substituting the PST--action
$S_0[A,C,a]$ with the dual PST--action $\widetilde S_0[A,C,b]$.
To see that this is the case one has to choose in 
$\widetilde S_U$
\bea
\label{gauge}
b_{\mu\nu}^{(0)}(x)&=&\ve_{\mu\nu\rho\sigma}n^\rho x^\sigma \quad 
\Rightarrow u^\mu=n^\mu\\
\nonumber
U^\mu(x)&=&n^\mu,
\eea
obtaining an action $\widetilde S_n[A,\Lambda,C,\Sigma,\varphi]$. Modulo
a suitable redefinition of $\Sigma$, one has
\bea
\widetilde S_n &=&
\int{1\over 2}\, dA\,{\cal Q}(n)\,dA+\left(dA-d\Lambda-{1\over 2}\Sigma\,ni_n
\right)\ve\,C \\ 
&\phantom{=}&-\sum_r\int d^4x\,\bar\varphi_r\left(D^\mu(\Lambda)D_\mu(\Lambda)
+m_r^2\right)\varphi_r.
\eea
As above we integrate over $A,C$ and $\Sigma$
$$
e^{i\widetilde S[\Lambda,\varphi]}\equiv
\int\{{\cal D} C\}\{ {\cal D}A\}\{{\cal D}\Sigma\}\,e^{i
\widetilde S_n[A,\Lambda,C,\Sigma,\varphi]}. 
$$
The integration over $C$ leads to the $\delta$--function
$\delta(dA-d\Lambda- {1\over 2}i_n n\Sigma)$, which implies 
$$
dA=d\Lambda+{1\over 2}\, i_n n\,\Sigma, \quad d(i_n n\,\Sigma)=0.
$$
Substituting these relations above and taking into account that
${\cal Q}(n) i_n n\,\Sigma ={1\over 2}\ve\, i_nn\,\Sigma$, the integration
over $A$ and $\Sigma$ becomes trivial. The result is
$$
\widetilde S[\Lambda,\varphi]=S_{ZW}[\Lambda,\varphi],
$$
i.e. again the Zwanziger action.

The connections presented above and in the preceding subsection
ensure that the quantum field theories based
on $S_U$, $\widetilde S_U$ and $S_{ZW}$ are equivalent. The PST approach
introduces additional non propagating auxiliary fields, but avoids some 
unpleasant features of the Zwanziger approach. Due to the presence
of the vector $n$ in the last one, the 
consistency check of the quantum theory requires the proof of 
Lorentz--invariance of the correlators of observables.  Therefore, the 
vector $n$ acquires only a posteriori the meaning of the direction of
the Dirac--string.  The entanglement of these two issues --
Lorentz--invariance and Dirac--string dependence -- in the
Zwanziger theory, leads, already at the level of point particles, to a rather 
complicated realization of Lorentz--invariance.  For point particles 
Zwanziger's action becomes 
\beq
\label{75a}
S_{ZW}^{pp}=
{1\over 2}\int\,  d\Lambda\,{\cal Q}(n)\,d\Lambda 
-\sum_r\,e_r\,\ve \int_{\gamma_r} \Lambda
-\sum_{r=1}^N m_r\int_{\gamma_r}d\tau_r.
\eeq
It can be obtained from the corresponding dual PST--action for
dynamical dyons, setting $u=n$ and choosing $C$'s satisfying $i_nC=0$.
A Lorentz transformation in \eref{75a} amounts then to make the replacement 
$n^\mu\rightarrow 
{n'}^\mu=\Lambda^\mu{}_\nu n^\nu$. From the equivalence with the dual 
PST--action it is clear that, to recover Lorentz--invariance, 
one has to modify the standard Lorentz--transformation law of $A$ as follows. 
First one
has to add a transformation of the type \eref{75}, which re-rotates the 
vector $n$ (the transformation of $b$ allows, indeed, to reach an 
arbitrary $n$). Second, since Zwanziger's action corresponds to a choice 
for the $C$'s restricted to $i_nC=0$, a change of $n$ has to be compensated
by a string shift, i.e. by $\delta A=-H_{n,n'}$; therefore, under these
modified Lorentz--transformations Zwanziger's action changes by an integer 
multiple of $2\pi$, if \eref{1} holds. 

These rather complicated arrangements are 
avoided in the classical PST--approach, due to its manifest 
$SO(1,3)$--invariance. Moreover, once the classical 
(exponentiated) PST--action is checked to be Dirac--string independent,
the quantum string--independence follows almost immediately.

The relation between the PST--approach and \cite{IENGO1} is less clear,
since the authors of that paper renounce to construct a classical field
theory action.  
    
\subsection{Equivalence of formulations with one and two vector potentials}

In the subsection 4.1 we gave an action for the $Z_4$--theory, in terms of a 
single vector potential, see the expression for $I_0[A^2,J^1,C^2]$ 
in eq. \eref{35}. There we observed also that the action
$I_0-{1\over 2}\int C^1C^2$ could be used, instead of the
PST--action, for a formulation of the $SO(2)$--theory in terms of a single    
vector potential. 
 
Vice versa, the action $S_0+{1\over 2}\int C^1C^2$ (or
$\widetilde S_0+{1\over 2}\int C^1C^2$) could be used for a formulation
of the $Z_4$--theory in terms of a doublet of vector potentials.

  The question, which arises naturally, is if the formulations with
one and two vector potentials are directly related: the functional
integration over one of the two vector potentials could give the 
corresponding formulation in terms of a single vector potential.
Actually, we have
two formulations in terms of vector potential doublets, based on $S_0$ 
and $\widetilde S_0$ respectively, and it turns out that, in this 
respect, the two formulations 
behave indeed differently. The reason has been anticipated above. In the
PST--action $S_0$ each vector potential carries one of  
the two transverse photon degrees of freedom. Therefore, integrating out
one of the two vector potentials leads to a non--local action for
the remaining vector potential, even in the absence of currents, as
can be checked also directly. On the other hand, in the dual 
PST--approach each vector potential carries two transverse degrees
of freedom, which become identified at the level of the equations of 
motion.  

Therefore, integrating the dual 
PST--action, augmented by the term ${1\over 2}\int C^1C^2$, over $A^1$ and
$b$, one can expect to obtain the action $I_0[A^2,J^1,C^2]$ for the 
$Z_4$--theory (while the analogous calculation with the PST--action
can surely not lead to $I_0$). We define $\widehat I_0$ by
$$
e^{i\widehat I_0[A^2,C^1,C^2]}=
\int\{{\cal D} A^1\} \{{\cal D} b\} \,e^{i\left(
\widetilde S_0[A,C,b] +{1\over 2}\int C^1C^2\right)}. 
$$
The integration over $b$ needs the insertion of a gauge--fixing 
$\delta$--function $\delta(b(x)-b^{(0)}(x))$, and the integration over
$A^1$ requires only to fix its $U(1)$--invariance. The  
gaussian integral over $A^1$, for a fixed $b^{(0)}$, can be conveniently
evaluated through  an $S$--duality transformation and turns out to
be independent of $b^{(0)}$. For the details of this calculation we refer 
the reader to the appendix. 

The result is, indeed,
\beq
\label{75b}
\widehat I_0=I_0.
\eeq 
The disappearance of  $b^{(0)}$, and the result itself, are
somewhat surprising since there is no symmetry which, a priori, protects
$\widehat I_0$ from a $b^{(0)}$--dependence or from non--local terms. Clearly,
the result is in agreement with the fact that a further functional 
integration over $A^2$ leads to the effective action $\Gamma_0^{Z_4}$;
but this would have happened also if one had started with the PST--action.

\subsection{$\vartheta$--angles and $SO(2)$--symmetry}

We anticipated that in the $SO(2)$--theory $\vartheta$--angles are already
present, while in the $Z_4$--theory they are introduced adding
to the action a term proportional to $\int FF$. In the case of the 
$SO(2)$--theory such terms are forbidden by the PST--symmetries.

To see how $\vartheta$--angles arise, it suffices to search for the general
solution of the $SO(2)$--invariant Dirac--Zwanziger quantization
condition for the charge vectors $(e_r,g_r)$,
\beq
\label{76}
e_r g_s -e_s g_r =4\pi n_{rs}. 
\eeq
Since this relation is invariant under $Sl(2,R)$ -- a three parameter group
-- and not only under
$SO(2)$, one expects that the general solution is characterized by three
continuous parameters. To see that this is indeed the case we observe that,
with an $SO(2)$ rotation, one can eliminate the magnetic charge of say 
the first charge vector, $g_1=0$. This leads to 
\beq
\label{77}
g_r={4\pi\over e_1}\,m_r,
\eeq 
with $m_r$ integer for all magnetic charges. Substituting
this back in \eref{76} one gets
$$
e_r={n_{rs}\over m_s}\,e_1+{e_s\over m_s}\, m_r  
$$
for each $r$ and $s$. Choosing a particular $s$ such that $L\equiv m_s
\neq 0$, and defining $e_0=e_1/L$, one can write the general solution as
\bea
\nonumber
e_r&=&e_0\left(n_r+{\vartheta\over 2\pi}\, m_r\right)\\
g_r&=&{4\pi\over e_0}\,{m_r\over L},
\label{78}
\eea
where $m_r,n_r$ and $L$ are integers restricted to
\beq
\label{79}
n_rm_s-n_sm_r=L\cdot K_{rs};
\eeq
this means that the left hand side has to be an integer multiple of $L$.
Apart from an $SO(2)$--rotation, the relations \eref{78} and \eref{79} specify
the general solution of \eref{76}. They represent the counterpart to the
relations \eref{46} in the $Z_4$--theory, modified by a $\vartheta$--term.

The integer $L$ can, in general, not be eliminated by rescaling $e_0$ or 
$\vartheta$. (For example, the solution corresponding to $n_r=m_r=M$ for 
any $r$ and 
$L$ arbitrary, can not be traced back to the case $L=1$.) Nevertheless,
the physically relevant solutions correspond to $L=1$. In fact, it is 
sufficient to assume that there exists a particle, say the electron, with the 
minimal electric charge $e_0$ and with vanishing magnetic charge: $n_s=1$, 
$m_s=0$. Equation \eref{79} implies then that all $m_r$ are integer 
multiples of 
$L$, and this amounts to set in \eref{78} $L=1$, upon rescaling $m_r$ and
$\vartheta$. Henceforth we will restrict to this case.

Taking the initial $SO(2)$--rotation into account, these relevant solutions
can then be expressed as
\beq\label{80}
{e_r \choose g_r}=T(\alpha)\left(
\matrix{e_0& {\vartheta\over 2\pi}\,e_0\cr
              0&{4\pi\over e_0}\cr}\right){n_r\choose m_r}\equiv
              \Omega(\alpha,e_0,\vartheta)\,{n_r\choose 4\pi\, m_r}.
\eeq
Here $T(\alpha)$ is an $SO(2)$--matrix, and 
$$
\Omega(\alpha,e_0,\vartheta)=T(\alpha)\left(
\matrix{e_0& {\vartheta\over 2\pi}\,{e_0\over 4\pi}\cr
              0&{1\over e_0}\cr}\right)
$$
parametrizes a generic element of $Sl(2,R)$, as anticipated above. 

Two theories whose charge vectors differ only by the matrix $T(\alpha)$
are equivalent, because this difference can be eliminated by an 
$SO(2)$--redefinition of the vector potential doublets
(a consequence of the $SO(2)$--duality invariance of the theory). 
This implies also that
the angle $\alpha$ is a fictitious coupling constant, and that the matrix 
$T(\alpha)$ can be set equal to unity. Thus the geometric meaning 
of \eref{80} is that the space of coupling constants is represented by
the coset $Sl(2,R)\over SO(2)$, parametrized by $\vartheta$ and $e_0$.
  
Apart from $\vartheta$ and $e_0$, the charge vectors are characterized
also by the vectors of integers $(n_r,m_r)$, which until now have been
kept fixed. If we consider also $Sl(2,Z)$--transformations of these 
integer 
vectors, we can find more equivalence relations between different theories.
It is sufficient to concentrate on the two generators ${\cal T}$ and 
${\cal S}$ of $Sl(2,Z)$. They act as
\bea\nonumber
{\cal T}\,:\quad &&e_0'=e_0\\ \nonumber
                 &&\vartheta'=\vartheta +2\pi\\ \nonumber          
                 &&n_r'=n_r-m_r\\ \nonumber
                 &&m_r'=m_r  \\ \nonumber
     &&\phantom{=}\\ \nonumber
{\cal S}\,:\quad &&e_0'={4\pi\over e_0}\,
    \sqrt{1+\left({\vartheta e_0^2\over 8\pi^2}\right)^2}\\ \nonumber           
                 &&\vartheta'=-{\left({e_0^2\over 4\pi}\right)^2\over  
1+\left({\vartheta e_0^2\over 8\pi^2}\right)^2}\, \vartheta\ \\ \nonumber
                 &&n_r'=m_r\\
                 &&m_r'=-n_r. \nonumber
\eea
Under a ${\cal T}$-transformation we have simply $(e_r,g_r)=(e_r',g_r')$
and the corresponding theories are trivially equivalent. Under ${\cal S}$
we have instead 
$$
 {e_r' \choose g_r'}=T(\alpha){e_r \choose g_r},
$$
where $T(\alpha)$ is an $SO(2)$--matrix, with rotation angle given by 
$$
{\rm tg}\,\alpha=-{8\pi^2\over\vartheta e_0^2}.
$$
Due to $SO(2)$--invariance the corresponding theories are then again 
equivalent. 

Notice, however, that the theory is {\it not invariant} under the shift
$\vartheta\rightarrow\vartheta+2\pi$ alone, because this shift amounts to 
$n_r\rightarrow n_r+m_r$, and this changes the electric charges of the
charge vectors. 
 
\section{Concluding remarks}

In this concluding section we mention briefly  some other possible 
applications and natural extensions of the methods developed in this paper.

The monopoles considered in this paper are Dirac--monopoles -- exhibiting a 
singularity at the monopole location -- and are typical for abelian gauge 
theories. A priori, the formalism developed in the paper does not apply to the 
non--singular `t Hooft--Polyakov monopoles \cite{HP}. They appear in the 
semi--classical analysis of non--abelian gauge theories whose gauge group is 
spontaneously broken to an abelian subgroup, containing at least one 
$U(1)$--factor; a prototype is given by the Georgi--Glashow model.
A way to exhibit explicitly `t Hooft--Polyakov monopoles in 
correlation functions of observables of the full quantum theory has been 
suggested by 't Hooft \cite{H}, and, following ideas 
similar to those outlined in sect 6.2.6, a
recipe to construct monopole correlation functions  has been proposed 
in \cite{PAM}.

Although in the semi--classical analysis these monopoles are non--singular,
it is well known \cite{AF} that they are mapped to Dirac--like monopoles in 
suitable unitary gauges and, therefore, in these gauges, the formalism  
developed in the present paper applies also to them. 

By adding to the action of e.g. the Georgi--Glashow model the topological 
$\vartheta$--term
\beq\label{teta}
{e^2 \vartheta \over  16\pi^2} \int {\rm Tr}\, ({\cal F} {\cal F}),
\eeq
where ${\cal F}$ is the non--abelian field strength two--form, the 
`t Hooft--Polyakov 
monopoles are turned into dyons. The integral appearing in \eref{teta} is a 
topological invariant, the second Chern class, and it takes values in the 
integers. Therefore, the quantum field theory with $\vartheta$--term is 
invariant under the shift $\vartheta \rightarrow \vartheta + 2 \pi$, in 
contrast with the remarks made at the end of the previous section for the 
abelian models considered in this paper. The different behaviour is related 
to the appearance of an {\it infinite} number of species of 
dyons in the Georgi--Glashow model, as discussed in \cite{H}. 

In abelian 
models invariance under $2 \pi$--shifts of $\vartheta$ can be achieved 
in the limiting case, where the integral $e^2_0/8 \pi^2\int F F$ is forced to 
be an integer, as in the Cardy--Rabinovici model \cite{CR}.

One of the principal advantages of the formalism presented in this 
paper is the implementation of Lorentz--invariance as a manifest symmetry,
at each step, while the quantum consistency of the theory relies on the 
Dirac--string independence. From the latter point of view we saw
that the classical point--particle theory contains already the relevant
informations, like the quantization conditions for the charges, the 
distinction
between $SO(2)$--theories and $Z_4$--theories, the effect of 
$\vartheta$-angles, and the current--current effective actions. 

At this ``semi--classical" level the formalism 
allows immediate extensions from point--particles (0--branes) to
arbitrary $p$--branes, coupled to $(p+1)$--form gauge potentials,
and to arbitrary dimensions. It can be applied
not only to $p$--brane dyons, living in $D=2p+4$ dimensions, but also
to a system of dual branes, for example strings and five--branes in 
ten dimensions. Results in this direction will be reported elsewhere 
\cite{LM}. 

\section{Appendix}

\subsection{Useful identities}

We collect here the basic identities, involving
differential $p$--forms, the $*$--operator and the interior product $i_U$ 
with a vector field $U^\mu$, used in the paper. The one--form $U$ 
is given by $U=dx^\mu U_\mu$. 
\bea\nonumber
d\left(\Phi_p\Phi_q\right)&=&\Phi_p d\Phi_q+(-)^q(d\Phi_p)\Phi_q\\
\nonumber
i_U\left(\Phi_p\Phi_q\right)&=&\Phi_p i_U\Phi_q+(-)^q(i_U\Phi_p)\Phi_q\\
\nonumber
d^2=&0&=(i_U)^2\\ \nonumber
i_U&=&(-)^p*U*\\ \nonumber
U&=&(-)^{p+1}*i_U*\\ \nonumber
{\phantom{x}}* i_U&=&U*\\ \nonumber
i_U*&=&-*U\\ \nonumber
\delta&=& *d*\\ \nonumber
{\phantom{x}} *^2&=&(-)^{p+1}\\ \nonumber
1&=&{1\over U^2}\left((-)^{p+1}Ui_U+*Ui_U*\right)\\ \nonumber
{\phantom{x}}*&=& {(-)^{p+1}\over U^2}\left(*Ui_U+Ui_U*\right)\\ \nonumber
\qua&=& d\delta+\delta d.\nonumber
\eea
For constant vector fields $n$ we have also 
$$
di_n +i_nd =\pa_n=n^\mu\pa_\mu.
$$       

\subsection{Proof of \eref{75b}} 

We begin by writing for $\widetilde S_0[A,C,b]$ an expression analogous
to \eref{ident}, keeping the notation used for that formula.
\beq
\label{ap1}
\widetilde S_0[A,C,b]=
\Gamma_0[C]+{1\over 2}\int (G+D)\,\widetilde\Omega(u) (G+D).
\eeq
The unique difference w.r.t. \eref{ident} lies in the operator
$\widetilde\Omega(u)$ which is here
$$
\widetilde\Omega(u) = {d*d\over\quadratello}-i_u*i_u.
$$
We remember that $G=dA^1-*dA^2$ and $D=C^1-*C^2$. The above identity
shows, by the way, that the effective action associated to $\widetilde S_0$
is $\Gamma_0[C]$.

Since we want now to integrate only over $A^1$ (and $b$) it is convenient 
to define
\bea\label{ap2}
\widehat A^2&=&A^2-*{d\over\qua}\,D\\
\widehat A^1&=& A^1 +{\delta\over\qua}\, D \nonumber,
\eea
such that $G+D=d\widehat A^1-*d\widehat A^2$. In terms of $\widehat A^I$ 
one can then write 
$$
\widetilde S_0[A,C,b]=\widetilde S_0[\widehat A,0,b]+\Gamma_0[C],
$$
where $\widetilde S_0[A,0,b]$ is the free dual PST--action, i.e. with 
vanishing currents, as can be seen by setting in \eref{ap1} $C=0$.
This means that
$$
\widetilde S_0[A,C,b]= {1\over 2}\int d\widehat A
\,{\cal Q}(u)\,d\widehat A  +\Gamma_0[C].
$$
This reduces the problem to the integration over $A^1$ and $b$ in the 
{\it free} dual PST--action,
\beq
\label{boh}
e^{iS[A^2]}\equiv \int \{ {\cal D}A^1\}\{ {\cal D}b\}\,e^{{i\over 2}\int
dA \,{\cal Q}(u)\,d A}\delta(b-b^{(0)}).
\eeq
To perform this integration it is convenient to perform an $S$--duality
of $A^1$, 
$$
e^{iS[A^2]}= e^{{i\over 2}\int dA^2 *ui_u dA^2}
\int \{ {\cal D}H\} \{ {\cal D}\Lambda \}\{ {\cal D}b\}
\,\delta(b-b^{(0)})\,
e^{i\int {1\over 2} H*ui_u H +(dA^2 ui_u-d\Lambda)H}.
$$          
Integration over $\Lambda$ gives, indeed, $dH=0\Rightarrow H=dA^1$.
The quasi--local symmetry of $A^2$, $\delta A^2=\chi^2, i_u d\chi^2=0$
has now to be accompanied by
\beq
\label{quasi}
\delta \Lambda=\chi^2.
\eeq
We perform now the shift of integration variable $H\rightarrow
H+ui_u*(dA^2-d\Lambda)$ which gives
\bea
e^{iS[A^2]}&=&e^{{i\over 2}\int dA^2 *dA^2}
\int \{ {\cal D}H\} \{ {\cal D}\Lambda \}\{ {\cal D}b\}
\,\delta(b-b^{(0)})\cdot\\
&\phantom{=}&
e^{i\int {1\over 2}\,H*ui_u H +{1\over 2}\, d\Lambda ui_u* d\Lambda
-d\Lambda ui_u* dA^2 - uHi_ud\Lambda}.
\eea
Since the  two orthogonal components of $H$, $ui_u H$ and $*ui_u*H$,
are now decoupled, the integration over these components factorizes and
gives    
$$
e^{iS[A^2]}=e^{{i\over 2}\int dA^2 *dA^2}
\int \{ {\cal D}\Lambda \}\{ {\cal D}b\}
\,\delta(b-b^{(0)})\,\delta(i_ud\Lambda)\,
e^{i\int {1\over 2}\, d\Lambda ui_u* d\Lambda
-d\Lambda ui_u* dA^2 }.
$$
But the $\delta$--function for $\Lambda$ means now that $d\Lambda$ 
is a pure gauge under \eref{quasi} and amounts, therefore, to
$\delta(d\Lambda)$. This implies that the integral over $\Lambda$
gives a constant, as does the final integration over $b$.
Thus, the result is $b^{(0)}$--independent and reads simply  
$$ 
S[A^2]={1\over 2}\int dA^2 * dA^2. 
$$
The functional integration over $A^1$ and $b$ of  
$exp\left(i\left(\widetilde S_0[A,C,b]+{1\over 2}\int C^1C^2 \right)\right)$
gives therefore 
$$
\widehat I_0={1\over 2}\int d\widehat A^2 * d\widehat A^2 +\Gamma_0[C]
+{1\over 2}\int C^1C^2. 
$$
Upon substituting \eref{ap2} and the explicit expression of 
$\Gamma_0[C]$, one obtains then easily
$$
\widehat I_0=\int {1\over 2}\,(dA^2+C^2)*(dA^2+C^2)-dA^2\,C^1=I_0,
$$ 
q.e.d.
\bigskip  
\paragraph{Acknowledgements.}

This work was supported by the 
European Commission TMR programme ERBFMPX-CT96-0045.




\vskip1truecm

\begin{thebibliography}{99}

\bibitem{SCHW1} J. Schwinger, Phys. Rev. {\bf 144} 4 (1966) 1087.

\bibitem{SCHW2} J. Schwinger, Phys. Rev. {\bf D12} 10 (1975) 3105.

\bibitem{ZW1} D. Zwanziger, Phys. Rev. {\bf D3} 4 (1971) 880.
 
\bibitem{ZW2} R.A. Brandt, F. Neri and D. Zwanziger, Phys. Rev. {\bf D19}
              4 (1979) 1153.
 
\bibitem{SCHWARZ} J.H. Schwarz and A. Sen, Phys. Lett. {\bf B312} (1993)
                  105; Nucl. Phys. {\bf B411} (1994) 35. 

\bibitem{DIRAC} P.A.M. Dirac, Proc. R. Soc. London {\bf A133} (1931) 60. 

\bibitem{PST}  P. Pasti, D. Sorokin and M. Tonin, Phys. Lett. 
{\bf B352} (1995) 59; P. Pasti, D. Sorokin and M. Tonin, Phys. Rev. 
{\bf D52R} (1995) 4277.

\bibitem{BERMED} N. Berkovits and R. Medina, Phys. Rev. {\bf D56} (1997) 
                 6388.

\bibitem{IENGO1} G. Calucci, R. Iengo and M.T. Vallon, Nucl. Phys. 
                {\bf B197} (1982) 93; {\bf B211} (1983) 77.

\bibitem{PSTCHIR} P. Pasti, D. Sorokin and M. Tonin, Phys. Rev. {\bf 
                 D55} (1997) 6292; Leuven Notes in 
Mathematical and Theoretical Physics (Leuven University Press), 
Series {\bf BV6} (1996) 167, hep--th/9509053.

\bibitem{KL} K. Lechner, Nucl. Phys. {\bf B537} (1999) 361.

\bibitem{IENGO2} G. Calucci, R. Iengo and M.T. Vallon, Nucl. Phys. 
                {\bf B223} (1983) 501.

\bibitem{BN} R.A. Brandt and F. Neri, Phys. Rev. {\bf D18} (1978) 2080.

\bibitem{deRham} G. de Rham, ``Differentiable manifolds. Forms, 
              Currents, Harmonic Forms", Springer-Verlag (1984). 
               
\bibitem{Diracveto} P.A.M. Dirac, Phys. Rev. {\bf 74} (1948) 817.

\bibitem{Feynman} R.P. Feynman, Rev. Mod. Phys. {\bf 20} (1948) 267;
                   Phys. Rev. {\bf 80} (1950) 440. 

\bibitem{path} T. Jaroszewicz and P.S. Kurzepa, Ann. Phys. {\bf 210}
               (1991) 255.             

\bibitem{Mandel} S. Mandelstam, Ann. Phys. {\bf 19} (1962) 1.

\bibitem{PAMdirac} P.A.M. Dirac, Can. J. Phys. {\bf 33} (1955) 650.

\bibitem{MPS} A. Maznytsia, C.R. Preitschopf and D. Sorokin, Nucl.
              Phys. {\bf B539} (1999) 438.

\bibitem{PAM} J. Froehlich and P.A. Marchetti, ``Gauge invariant charged
              monopole and dyon fields in gauge theories",
              hep-th/9812004, to appear in Nucl. Phys. B.

\bibitem{Witten} E. Witten, Phys. Lett. {\bf B86} (1979) 283.

\bibitem{GJ} J. Glimm and A. Jaffe, ``Quantum Physics. A Functional Integral 
              Point of View", Springer-Verlag (1981).

\bibitem{St} F. Strocchi, Commun. Math. Phys. {\bf 56} (1977) 57.  

\bibitem{El} S. Elitzur, Phys. Rev. {\bf D12} (1975) 3978; 
             G.F. De Angelis, D. De Falco and F. Guerra, Phys. 
             Rev. {\bf D17} (1978) 1624.

\bibitem{FP} J. Froehlich and P.A. Marchetti in ``The Algebraic Theory of 
             Superselection Sectors. Introduction and Recent Results", 
             D. Kastler ed., World Scientific 1990; Europhys. Lett. 
             {\bf 2} (1986) 933.

\bibitem{HP} G. `t Hooft, Nucl. Phys. {\bf B79} (1974) 276; A.M. Polyakov, 
             JETP Lett. {\bf 20} (1974) 194.

\bibitem{H}  G. `t Hooft, Nucl. Phys. {\bf B190} (1981) 455; see also A.S. 
             Kronfeld, M. L. Laursen, G. Schierholz and U.J. Wiese, Phys. 
             Lett. {\bf B198} (1987) 516.

\bibitem{AF} J. Arafune, P.G.O. Freund and C.J. Goebel, J. Math. Phys. 
             {\bf 16} (1975) 433.

\bibitem{P}  P.A. Marchetti, ``Bosonization in condensed matter systems", 
             Chia, Italy, 1995, hep-th/9511100.

\bibitem{CR} J.L. Cardy and E. Rabinovici, Nucl. Phys. {\bf B205} (1982)
             1.

\bibitem{LM} K. Lechner and P.A. Marchetti, in preparation.
\end{thebibliography}
\end{document}